\documentclass[aps,prd,floats,floatfix,nofootinbib]{revtex4-1}
\usepackage{graphicx,amsmath}
\usepackage{amssymb}
\usepackage{amsfonts}
\usepackage{hyperref}
\usepackage[bottom]{footmisc}

\begin{document}

\title{Discovering Strong Top Dynamics at the LHC}

\author{R.\ Sekhar Chivukula}
\email {sekhar@msu.edu}
\affiliation {Department of Physics and Astronomy,
Michigan State University,
East Lansing, MI 48824, USA}

\author{Baradhwaj Coleppa}
\email {barath@physics.carleton.ca }
\affiliation {Ottawa-Carleton Institute for Physics,
Carleton University,
Ottawa, Ontario K1S 5B6, Canada}

\author{Pawin Ittisamai}
\email {ittisama@pa.msu.edu}
\affiliation {Department of Physics and Astronomy,
Michigan State University,
East Lansing, MI 48824, USA}

\author{Heather E.\ Logan}
\email {logan@physics.carleton.ca }
\affiliation {Ottawa-Carleton Institute for Physics,
Carleton University,
Ottawa, Ontario K1S 5B6, Canada}

\author{Adam Martin}
\email {aomartin@fnal.gov }
\affiliation {Theoretical Physics Department,
Fermilab,
Batavia, IL 60510, USA}

\author{Jing Ren}
\email {jingren@pa.msu.edu}
\affiliation {Department of Physics and Astronomy,
Michigan State University,
East Lansing, MI 48824, USA}
\affiliation { Center for High Energy Physics and Institute of Modern Physics, 
Tsinghua University, 
Beijing 100084, China.}

\author{Elizabeth H.\ Simmons}
\email {esimmons@pa.msu.edu}
\affiliation {Department of Physics and Astronomy,
Michigan State University,
East Lansing, MI 48824, USA}

\date{\today}

\begin{abstract}
We analyze the phenomenology of the top-pion and top-Higgs states in models with strong top dynamics, 
and translate the present LHC searches for the Standard Model  Higgs into bounds on these scalar states. 
We explore the possibility that the new state at a mass of approximately 125 GeV
observed at the LHC is consistent with a neutral pseudoscalar top-pion state. We demonstrate that
a neutral pseudoscalar top-pion can generate the diphoton signal at the observed rate.  However, the region of model parameter space where this is the case does not correspond to classic topcolor-assisted technicolor scenarios with degenerate charged and neutral top-pions and a top-Higgs mass of order $2m_t$; rather, additional isospin violation would need to be present and the top dynamics would be more akin to that in top seesaw models.  Moreover, the interpretation of the new state as a top-pion can be sustained only if the $ZZ$ (four-lepton) and $WW$ (two-lepton plus missing energy) signatures initially observed at the 3$\sigma$ level decline in significance as additional data is accrued.

\end{abstract}

\maketitle

\section{Introduction}

The heavy mass of the top-quark necessarily implies that it couples more strongly to the electroweak symmetry
breaking sector than any other quark or lepton, and suggests that the top-quark itself may play a role in electroweak symmetry breaking \cite{Miransky:1988xi,Miransky:1989ds,Nambu:1989jt,Marciano:1989xd,Bardeen:1989ds}. The Top Triangle Moose model \cite{Chivukula:2009ck} is a consistent low-energy effective theory for models with separate sectors for dynamically generating the masses of the top quark and the weak vector bosons.  It can be used to investigate the phenomenology of a wide range of theories that include new strong top quark dynamics \cite{Hill:1991at,Hill:1994hp,Lane:1995gw,Lane:1996ua,Popovic:1998vb, Hill and Simmons, Braam:2007pm}. In previous work
\cite{Chivukula:2011ag,Chivukula:2011dg} the authors have investigated the phenomenology of the scalar sector
of the top-triangle model, and have explored the constraints placed on the ``top-Higgs" boson present in these
models by searches at the LHC for a standard model Higgs boson.  We concluded that the top-Higgs boson mass was constrained to lie above 300 GeV for the region of Top Triangle Moose parameter space corresponding to numerous strong dynamics  models. 

In this work we update our results
on top-Higgs searches in light of new data from the LHC and consider bounds on the ``top-pions" that are also present.
In particular, we explore the possibility that the new boson with a mass of approximately 125 GeV \cite{CMSgaga,CMS:2012gu,ATLASgaga,ATLAS:2012gk} observed at the LHC is consistent with a neutral pseudoscalar top-pion state.\footnote{The possibility that the boson observed at the LHC is a pseudo-scalar has been considered by a number of authors recently \protect\cite{Bernreuther:2010uw,Burdman:2011ki,Holdom:2012pw,Cervero:2012cx,Frandsen:2012rj,Moffat:2012ix,Barroso:2012wz}.} We demonstrate that
a neutral pseudoscalar top-pion can generate the diphoton signal at the observed rate.  However, the region of model parameter space where this is the case does not correspond to classic topcolor-assisted technicolor scenarios with degenerate charged and neutral top-pions and a top-Higgs mass of order $2m_t$; rather, additional isospin violation would need to be present and the top dynamics would be more akin to that in top seesaw models \cite{Dobrescu:1997nm,Chivukula:1998wd,He:2001fz}.  Moreover, the interpretation of the new state as a top-pion can be sustained only if the $ZZ$ (four-lepton) and $WW$ (two-lepton plus missing energy) signatures initially observed at the 3$\sigma$ level decline in significance as additional data is accrued.  

On one level, the Top Triangle Moose model is an example of a deconstructed Higgsless model of electroweak symmetry breaking. Inspired by the possibility of maintaining perturbative unitarity in extra-dimensional models through heavy vector resonance exchanges in lieu of a Higgs \cite{SekharChivukula:2001hz,Chivukula:2002ej,Chivukula:2003kq}, Higgsless models were initially introduced in an extra-dimensional context as $SU(2)\times SU(2)\times U(1)$ gauge theories living in a slice of $AdS_5$, with symmetry breaking codified in the boundary condition of the gauge fields \cite{Csaki:2003dt,Csaki:2009bb,Cacciapaglia:2006gp,Cacciapaglia:2004rb,Cacciapaglia:2004jz,Csaki:2003zu}. The low energy dynamics of these extra-dimensional models can be understood in terms of a collection of 4-D theories, using the principle of ``deconstruction'' \cite{Hill:2000mu,ArkaniHamed:2001ca}. Essentially, this involves latticizing the extra dimension, associating a 4-D gauge group with each lattice point and  connecting them to one another by means of nonlinear sigma models; the picture that emerges is called a ``Moose'' diagram~\cite{Georgi:1985hf}. The five dimensional gauge field is now spread in this theory as four dimensional gauge fields residing at each lattice point, and the fifth scalar component residing as the eaten pion in the sigma fields.    

 The key features of these models \cite{Casalbuoni:1985kq,Casalbuoni:1995qt,Chivukula:2005ji,Chivukula:2005xm,Chivukula:2005bn,Kurachi:2004rj,Chivukula:2004af,Chivukula:2004pk,Lane:2009ct,Foadi:2003xa,Foadi:2004ps,Foadi:2005hz} that are relevant to our discussion are  as follows:  Spin-1 resonances created by the strong dynamics underlying the sigma fields are described as massive gauge bosons, following the Hidden-Local-Symmetry scenario originally used for QCD~\cite{Bando:1987br,Bando:1985rf,Bando:1984pw,Bando:1984ej,Bando:1987ym} and also the BESS \cite{BESS-1,BESS-2} models.  The phenomenology of those resonances in the Top Triangle Moose have been discussed in Refs.   \cite{Chivukula:2006cg,Chivukula:2009ck}.  Standard model (SM) fermions reside primarily on the exterior sites -- the sites approximately corresponding to $SU(2)_w$ and $U(1)_Y$ gauge groups; these fermions become massive through mixing with massive, vector-like fermions located on the interior, `hidden' sites.  The phenomenology of these fermions has previously been discussed in \cite{Chivukula:2009ck,Chivukula:2011ag}.  Precision electroweak parameters~\cite{Peskin:1991sw}, are accommodated by adjusting the SM fermion's distribution across sites \cite{Foadi:2004ps} to match the gauge boson distribution, a process called ``ideal delocalization'' \cite{Chivukula:2005xm}. This is identical to the solution used in extra-dimensional Higgsless models, where the spreading of a fermion among sites becomes a continuous distribution, or profile, in the extra dimension~\cite{Cacciapaglia:2004rb}.  
  
The $AdS/CFT$ correspondence suggests that these weakly-coupled Higgsless models can be understood to be dual to the strongly coupled models of electroweak symmetry breaking.  Indeed the Top Triangle Moose is a deconstructed analog of topcolor-assisted technicolor (TC2) \cite{Hill:1991at,Hill:1994hp,Lane:1995gw,Lane:1996ua,Popovic:1998vb, Hill and Simmons, Braam:2007pm}, a scenario of dynamical electroweak symmetry breaking in which the new
strong dynamics is partitioned into two different sectors. The technicolor sector \cite{Weinberg:1979bn,Susskind:1978ms} is responsible for the bulk of electroweak
symmetry breaking, through condensation of a technifermion bilinear, and is therefore characterized by a scale $F \sim
v$, where $v=$ 246 GeV is the EWSB scale. Consequently, technicolor dynamics is responsible for the majority
of the weak gauge boson masses and, more indirectly  \cite{Eichten:1979ah,Dimopoulos:1979es}, the masses of the light fermions. 
The second strong sector, the topcolor sector \cite{Hill:1991at,Hill:1994hp}, communicates
directly with the top quark. Its purpose is to generate a large
mass for the top quark through new strong dynamics that cause top quark condensation \cite{Miransky:1988xi,Miransky:1989ds,Nambu:1989jt,Marciano:1989xd,Bardeen:1989ds}. In generating a top-quark mass, this second sector also helps to break the electroweak symmetry. If the characteristic scale of the
topcolor sector is low, $f \ll F$, it plays only a minor role in
electroweak breaking, but can still generate a sufficiently large top-quark mass
given a strong enough top-topcolor coupling. Because electroweak symmetry is effectively broken twice in this
scenario, there are two sets of Goldstone bosons. One linear combination of the weak-triplet Goldstone bosons (the combination primarily composed of technifermions)
 is eaten to become the longitudinal modes
of the $W^{\pm}/Z^0$, while the orthogonal triplet and accompanying weak singlet state remain in the
spectrum. These remaining states, typically referred to as the top-pions and the top-Higgs, are the focus of this paper.

The LHC evidence for a new boson is composed of several components, based on separate event samples optimized to be sensitive
to the production of the new boson via gluon-fusion, via vector-boson fusion, or in association with an electroweak 
boson or a top-quark pair, and the subsequent decay of the boson to two photons, two massive electroweak bosons, or
 pairs of tau-leptons or bottom-quarks \cite{CMS:2012gu,ATLAS:2012gk}. While the totality of evidence including all 
 subchannels provides convincing evidence of a new bosonic state -- one consistent with a Standard Model (SM) Higgs -- the statistical 
 significance of the different subchannels varies, and it is not yet certain that the object discovered is {\it the} Higgs
 boson. With the current data the evidence for the new boson is strongest in the diphoton channel, with
 a local $p$-value showing that the ``background-only" hypothesis is excluded at more than the 4$\sigma$ level by both
 experiments (a level which is {\it larger} than would have been expected with the current data set for the SM Higgs). The evidence in the next
 most sensitive decay channel, $ZZ^*$ subsequently decaying to four charged-leptons ($e$ or $\mu$), is also strong -- with a local
 $p$-value rejecting the background-only hypothesis at the 3$\sigma$ level. The search for the $WW^*$ decay mode, in which the
 $W$-bosons subsequently decay to $e$ or $\mu$ and corresponding neutrino, is less constraining since it is not possible
 to measure the diboson invariant mass -- though the background only hypothesis is disfavored by 2-3 $\sigma$. Finally, the evidence for the decay of the new boson to fermions, either tau-leptons or bottom-quarks, is so far inconclusive.

Our goal in this paper is to further the phenomenological investigation of the top-pions and top-Higgs at the LHC that was started in \cite{Chivukula:2009ck,Chivukula:2011ag,Chivukula:2011dg}.
We begin in Sec.~\ref{sec:model} by setting out the relevant details of
the Top Triangle Moose model.   Sections \ref{sec:neutralpionlimits} -- \ref{sec:tophiggslimits} contain the bulk of our
phenomenological results.   In Sec.~\ref{sec:neutralpionlimits} we first consider the possibility that the diphoton signal
observed at the LHC arises from the neutral pseudoscalar top-pion and find the range of model parameters
consistent with these experimental results. Since this object is a psuedoscalar, it lacks
tree-level couplings to $ZZ$ and $WW$ \cite{Frandsen:2012rj,Low:2012rj,Coleppa:2012eh}.  While
the top-pion can decay to $ZZ$ or $WW$ through a top-quark loop, we show that these effects would be too
small to be observable in the current data. In Sec.~\ref{sec:chargedpionlimits} we demonstrate that, for
the value of model parameters such that the neutral top-pion can account for the observed LHC diphoton
signal, the properties of top-quark decay imply that the corresponding charged top-pions would have to be heavier
than 150 GeV. As reviewed in Appendix \ref{appendix:NJL}, however, this implies that the model would need to
include more isospin violation than is the minimum required to produce a heavy top-quark -- i.e., 
more isospin violation than is usually assumed to exist in these models. In Sec.~\ref{sec:tophiggslimits}
we review and update the constraints previously derived in \cite{Chivukula:2011dg}, in the case that
the 125 GeV object is associated with the neutral top-pion.
 We summarize our findings and discuss their implications in 
Sec.~\ref{sec:conclusion}. 

\section{The Scalar Spectrum and Properties}
\label{sec:model}

Probing the dynamics of topcolor assisted technicolor will involve discovering
the top-Higgs and top-pions which are associated with the generation of the large top-quark mass, and measuring
their properties. In this section we describe briefly our expectations for the properties of these
states, and summarize the model-dependence of their couplings.

\subsection{The Triangle Moose Model}

\begin{figure}[h!]
\begin{center}
\includegraphics[angle=0,width=1.5in]{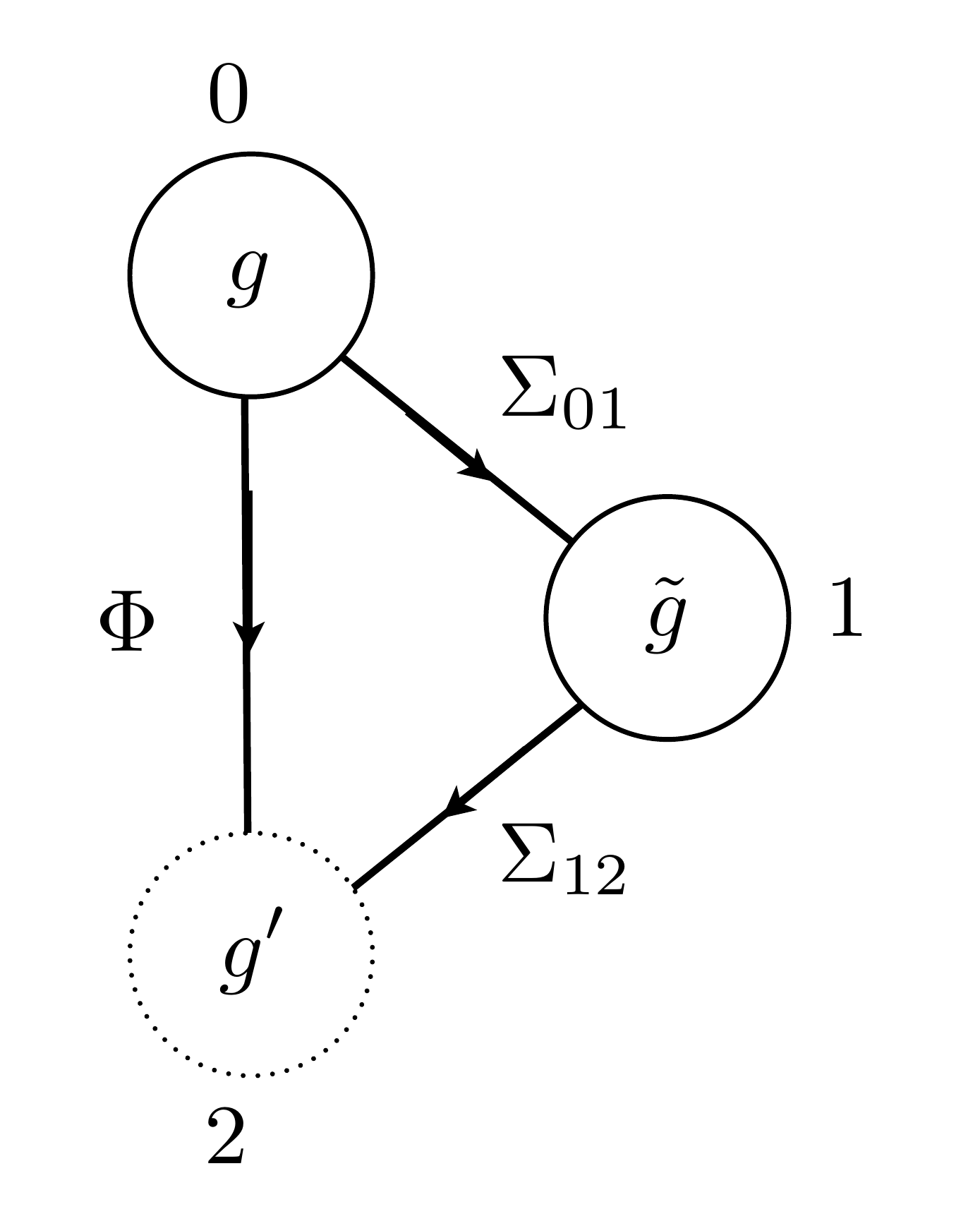}
\caption{The gauge structure of the model in Moose notation. $g$ and
  $g^{\prime}$ are approximately the Standard Model $SU(2)$ and hypercharge gauge
  couplings while $\tilde{g}$ represents the `bulk' gauge
  coupling. The left (right) handed light fermions are mostly
  localized at site 0 (2) while their heavy counterparts are mostly at
  site 1. The links connecting sites 0 and 1 and sites 1 and 2 are nonlinear sigma model fields while the one connecting sites 0 and 2 is
  a linear sigma field. Site 2 is dotted to indicate that only the $\tau_3$ component is gauged.}
\label{fig:Triangle}
\end{center}
\end{figure}

The Top Triangle Moose model~\cite{Chivukula:2009ck} is shown in Moose
notation in Fig.~\ref{fig:Triangle}.  The circles represent global
SU(2) symmetry groups; the full SU(2) at sites 0 and 1 are gauged with
gauge couplings $g$ and $\tilde g$, respectively, while the $\tau^3$
generator of the global SU(2) at site 2 is gauged with U(1) gauge
coupling $g^{\prime}$.  The lines represent spin-zero link fields
which transform as a fundamental (anti-fundamental) representation of
the group at the tail (head) of the link.  $\Sigma_{01}$ and
$\Sigma_{12}$ are nonlinear sigma model fields, describing the
technicolor/three-site \cite{Chivukula:2006cg} sector of the theory, while $\Phi$ (the
top-Higgs doublet) is a linear sigma model field arising from top-color \cite{Hill:1991at,Hill:1994hp}. 

The kinetic energy terms of the link fields corresponding to these
charge assignments are:
\begin{equation}
   \mathcal{L}_{gauge}=
   \frac{F^{2}}{4} 
   \textrm{Tr}[(D_{\mu}\Sigma_{01})^{\dagger}D^{\mu}\Sigma_{01}]
   + \frac{F^{2}}{4}
   \textrm{Tr}[(D_{\mu}\Sigma_{12})^{\dagger}D^{\mu}\Sigma_{12}]
   + (D_{\mu}\Phi)^{\dagger}D^{\mu}\Phi,
\label{eqn:Gauge L}
\end{equation}
where the covariant derivatives are:
\begin{eqnarray}
   D_{\mu}\Sigma_{01} &=& \partial_{\mu}\Sigma_{01} + igW_{0\mu} \Sigma_{01}
   - i\tilde{g}\Sigma_{01}W_{1\mu}, \nonumber \\
   D_{\mu}\Sigma_{12} &=& \partial_{\mu}\Sigma_{12} 
   + i\tilde{g}W_{1\mu} \Sigma_{12} 
   - ig^{\prime} \Sigma_{12}\tau^{3}B_{\mu}, \nonumber \\
   D_{\mu}\Phi &=& \partial_{\mu}\Phi + igW_{0\mu} \Phi
   - \frac{ig^{\prime}}{2}B_{\mu}\Phi.
\label{eqn:covariant}
\end{eqnarray}
Here the gauge fields are represented\,\footnote{Here the subscripts appearing in
  the fields will refer to the ``site'' numbers and the superscripts
  will be reserved for $SU(2)$ indices.} by the matrices $W_{0\mu} =
W_{0\mu}^{a}\tau^{a}$ and $W_{1\mu} = W_{1\mu}^{a}\tau^{a}$, where
$\tau^a = \sigma^a/2$ are the generators of SU(2).  The nonlinear
sigma model fields $\Sigma_{01}$ and $\Sigma_{12}$ are 2$\times$2
special unitary matrix fields.  To mimic the symmetry breaking caused by
underlying technicolor and topcolor dynamics, we assume all link
fields develop vacuum expectation values (vevs):
\begin{equation}
   \langle \Sigma_{01} \rangle = \langle \Sigma_{12} \rangle 
   = \mathbf{1}_{2\times 2}, \qquad \qquad
   \langle \Phi \rangle = 
   \left( \begin{array}{c} f/\sqrt{2} \\ 0 \end{array} \right).
\end{equation}
In order to obtain the correct amplitude for muon decay, we parameterize the
vevs in terms of a new parameter $\omega$,
\begin{equation}
   F=\sqrt{2}\,v\,\cos\,\omega, \qquad \qquad 
   f=v\,\sin\,\omega,
   \label{eq:Ff}
\end{equation}
where $v = 246$~GeV is the weak scale.  We will explore the parameter range\footnote{The extreme case in which $\sin\omega \to 1$ would have a rather different phenomenology, as the properties of the top-Higgs boson would approach those of the Standard Model Higgs boson, the top-Higgs could potentially be light, and the top-pions would become heavier.} $0.2 \leq \sin\omega \leq 0.8$, in which the Top Triangle Moose acts as a low-energy effective theory for a variety of models with strong top dynamics \cite{Chivukula:2011dg}. As a consequence of the  vacuum expectation values, the gauge symmetry is broken all the way
down to electromagnetism and we are left with massive gauge bosons
(analogous to techni-resonances), top-pions and a top-Higgs. To keep track of
how the degrees of freedom are partitioned after we impose the
symmetry breaking, we expand $\Sigma_{01}$, $\Sigma_{12}$ and $\Phi$
around their vevs. The coset degrees of freedom in the bi-fundamental
link fields $\Sigma_{01}$ and $\Sigma_{12}$ can be described by
nonlinear sigma fields:
\begin{equation}
   \Sigma_{01}=\textrm{exp}\left(\frac{2i\pi_{0}^{a}\tau^{a}}{F}\right),
   \qquad \qquad 
   \Sigma_{12}=\textrm{exp}\left(\frac{2i\pi_{1}^{a}\tau^{a}}{F}\right),
\label{eq:vevs}
\end{equation}
while the degrees of freedom in $\Phi$ fill out a linear
representation,
\begin{equation}
   \Phi= \left( \begin{array}{c}
     (f + H_t + i\pi_{t}^{0})/\sqrt{2} \\
     i\pi^{-}_{t} \end{array} \right).
\label{eqn:phi representation}
\end{equation}

The gauge-kinetic terms in Eq.~\eqref{eqn:Gauge L} yield mass matrices
for the charged and neutral gauge bosons. 
The photon remains massless and is given by the exact expression
\begin{equation}
  A_{\mu} = \frac{e}{g} W_{0 \mu}^3 + \frac{e}{\tilde g} W_{1 \mu}^3
  + \frac{e}{g^{\prime}} B_{\mu},
\end{equation}
where $e$ is the electromagnetic coupling. Normalizing the photon eigenvector, we get the relation between the coupling constants:
\begin{equation}
 \frac{1}{e^2}=\frac{1}{g^2}+\frac{1}{\tilde{g}^2}+\frac{1}{g'^2}.
\end{equation}
This invites us to conveniently parametrize the gauge couplings in terms of $e$ by

\begin{equation}
  g = \frac{e}{\sin\theta \cos\phi} = \frac{g_0}{\cos\phi}, \qquad \qquad
  \tilde g = \frac{e}{\sin\theta \sin\phi} = \frac{g_0}{\sin\phi}, 
  \qquad \qquad
   g^{\prime} = \frac{e}{\cos\theta}.
\label{eqn:gauge couplings}
\end{equation}
We will take $\tilde g \gg g$, which implies that $\tan\phi \equiv x$
is a small parameter.

\subsection{The Triangle Moose Potential: Scalar Spectrum and Isospin Violation}

Counting the number of degrees of freedom, we see that there are six
scalar degrees of freedom on the technicolor side ($\Sigma_{01},
\Sigma_{12}$) and four on the topcolor side ($\Phi$). Six of these
will be eaten to form the longitudinal components of the $W^{\pm}$,
$Z^0$, $W^{\prime \pm}$, and $Z^{\prime 0}$.  This leaves one isospin
triplet of scalars, the top-pions $\Pi^a_t$, and the top-Higgs $H_t$ as physical states in the
spectrum.  While the interactions in Eq.~\eqref{eqn:Gauge L} are
sufficient to give mass to the gauge bosons, the top-pions and
top-Higgs remain massless at tree level. Quantum corrections will give
the top-pions a mass, however this loop-level mass is far too small to be consistent with experimental constraints. To generate phenomenologically
acceptable masses for the top-pions and top-Higgs, we add three\footnote{In \protect\cite{Chivukula:2009ck} the possibility of isospin violation,
and hence the last term in Eq. (\protect\ref{eq:Vhiggs}), was neglected. As we show in Appendix \protect\ref{appendix:NJL}, isospin violation in the top color sector is usually assumed to be small, and hence the size of the dimensionless parameter
$\epsilon$,  is {\it small}. We introduce it here to explore the phenomenology that would arise from
non-degenerate top-pions.}
additional interactions:
\begin{equation}
   \mathcal L_M =  - \lambda\, {\rm Tr} 
   \left( M^{\dagger}M - \frac{f^2}{2} \right)^2 
   - \kappa f^2\, {\rm Tr} 
   \left| M - \frac{f}{\sqrt 2} \Sigma_{01}\Sigma_{12} \right|^2
   + \left\{ \epsilon f^2 \left( {\rm Tr} \left[ M^{\dagger} \Sigma_{01} \Sigma_{12} \tau^3 \right] \right)^2 + {\rm h.c.} \right\},
\label{eq:Vhiggs}
\end{equation}
where the first of these interactions arises from topcolor interactions, the second from ETC-like interactions~\cite{Eichten:1979ah,Dimopoulos:1979es}, and the third is an example of possible isospin-violating interactions  in the top-color sector. 
Here $\lambda$, $\kappa$, and $\epsilon$ are three new dimensionless parameters that depend on
the details of the top-color dynamics, $f$ is the same
vacuum expectation value appearing in Eq.~\eqref{eq:Ff}, and $M$ is
the $\Phi$ field expressed as a matrix\footnote{This corrects the expression in \protect\cite{Chivukula:2009ck}.}, schematically given by 
$M = ( \Phi, \tilde \Phi)$ with $\tilde \Phi = -i \sigma_2 \Phi^*$:
\begin{equation}
   M = \left( \begin{array}{cc}
      (f + H_t + i \pi_t^0)/\sqrt{2}& i \pi_t^+ \\
     i \pi_t^-  & (f + H_t - i \pi_t^0)/\sqrt{2}\end{array} \right),
\end{equation}
where $\pi_t^+ = (\pi_t^-)^*$.  The first term in
Eq.~\eqref{eq:Vhiggs} depends only on the modulus of $M$, and
therefore contributes only to the mass of the top-Higgs. The second and third
terms give mass to both the top-Higgs and the physical (uneaten)
combination of pion fields, as we will show shortly.  Because these
masses depend on three parameters, $\lambda$, $\kappa$, and $\epsilon$, we can treat
the mass of the top-Higgs and the masses of the uneaten charged and neutral top-pions
as three independent parameters. In addition to generating masses, the
potential in Eq.~\eqref{eq:Vhiggs} also induces interactions between
the top-Higgs and top-pions which are important in our analysis.

The next step towards understanding top-pion phenomenology is to
identify the combination of degrees of freedom which make up the
physical (uneaten) top-pions.  While the top-Higgs $H_t$ remains a
mass eigenstate, the pions $\pi_0^a$, $\pi_1^a$ and $\pi_t^a$ mix.  We
can identify the physical top-pions as the linear combination of
states that cannot be gauged away. We do this by isolating the
Goldstone boson states that participate in interactions of the form
$V_{\mu} \partial^{\mu} \pi$ in the Lagrangian.
We start by expanding the nonlinear sigma fields to first order in
$\pi/F$,
\begin{eqnarray}
   \Sigma_{01} &=& 1 + \frac{2i\pi_{0}^{a}\tau^{a}}{F} 
   + \mathcal{O}\left(\frac{\pi^2}{F^2}\right),
   \label{eqn:sigma01} \\
   \Sigma_{12} &=& 1 + \frac{2i\pi_{1}^{a}\tau^{a}}{F}
   + \mathcal{O}\left(\frac{\pi^2}{F^2}\right).
\label{eqn:sigma12} 
\end{eqnarray} 
Plugging this in Eq.~\eqref{eqn:Gauge
  L}, we can read off the various interaction terms. 
The gauge-Goldstone mixing terms are of the form:
\begin{equation}
   \mathcal{L}_{\textrm{mixing}} =
   \frac{g}{2} W_0^{a\mu} \partial_{\mu} \left[F\pi_0^a + f\pi_t^a \right] 
   + \frac{\tilde{g}}{2} W_1^{a\mu} \partial_{\mu} 
   \left[F \pi_1^a - F \pi_0^a \right]
   - \frac{g^{\prime}}{2} B_2^{\mu} \partial_{\mu} 
   \left[F \pi_1^3 + f \pi_t^3 \right].
   \label{eq:gauge-Goldstone}
\end{equation}
Note that the pion combination in the third term can be written as a linear
combination of those appearing in the first two terms:
\begin{equation}
  F \pi_1^3 + f \pi_t^3 = [F \pi_0^3 + f \pi_t^3] + [F \pi_1^3 - F \pi_0^3].
\end{equation}
The two eaten triplets of pions span the linear combinations that
appear in the first two terms of Eq.~\eqref{eq:gauge-Goldstone},
leaving the third linear combination as the remaining physical
top-pions, which we will denote $\Pi_t^a$:
\begin{equation}
   \Pi_t^a = - \sin\omega \left(\frac{\pi_0^a + \pi_1^a}{\sqrt{2}} \right)
   + \cos\omega \, \pi_t^a,
\end{equation}
where we have normalized the state properly using the definitions
of $F$ and $f$ in Eq.~\eqref{eq:Ff}.

The physical top-pions can also be identified by expanding the top-Higgs
potential given in Eq.~\eqref{eq:Vhiggs} and collecting the mass terms.
The physical masses of the top-Higgs and top-pions are
\begin{eqnarray}
	M_{\Pi^\pm_t}^2 &=& 2 \kappa v^2 \tan^2 \omega \nonumber \\
	M_{\Pi^0_t}^2 &=& 2 (\kappa - \epsilon) v^2 \tan^2 \omega \nonumber \\
	M_{H_t}^2 &=& 2 (4 \lambda + \kappa) v^2 \sin^2\omega  = 8 \lambda v^2 \sin^2\omega + M^2_{\Pi^+_t}\cos^2\omega.
\label{eq:pimass}
\end{eqnarray}
while the other two linear combinations of pions are massless, as true Goldstone bosons should be.  Equation~(\ref{eq:Vhiggs}) also
contains trilinear couplings between $H_t$ and two top-pions; 
the Feynman rules for the $H_t \Pi^+_t \Pi^-_t$ and $H_t \Pi^0_t \Pi^0_t$ interactions are given by
\begin{eqnarray}
	H_t \Pi^+_t \Pi^-_t: \ && -2i v \sin\omega \left[ 4 \lambda \cos^2\omega + \kappa \frac{\sin^4\omega}{\cos^2\omega} \right] \nonumber \\
	&& = \frac{-i}{v \sin\omega} \left[ M_{H_t}^2 \cos^2 \omega - M_{\Pi^+_t}^2 + 2 M_{\Pi^+_t}^2 \sin^2 \omega \right] \nonumber \\
	H_t \Pi^0_t \Pi^0_t: \ &&  -2i v \sin\omega \left[ 4 \lambda \cos^2\omega + \kappa \frac{\sin^4\omega}{\cos^2\omega} - 2 \epsilon \frac{\sin^2\omega}{\cos^2\omega} \right] \nonumber \\
	&& = \frac{-i}{v \sin\omega} \left[ M_{H_t}^2 \cos^2 \omega - M_{\Pi^+_t}^2 + 2 M_{\Pi^0_t}^2 \sin^2 \omega \right].
\end{eqnarray}
These couplings are important for top-Higgs decays when $M_{H_t} > 2
M_{\Pi_t}$.  

For the purposes of our phenomenological analysis we will take the masses of
the top-Higgs, and of the charged and neutral top-pions as independent parameters. To give a sense of
what might be expected from TC2 dynamics, we have looked at the expectations for these parameters in a Nambu-- Jona-Lasinio (NJL) 
\cite{Nambu:1961tp}  approximation for the topcolor dynamics; our NJL
calculation is summarized in Appendix \ref{appendix:NJL}. From the NJL analysis we find:

\begin{itemize}

\item The top-Higgs mass satisfies $M_{H_t} ={\cal O}(2m_t)$ \cite{Nambu:1961tp}. This  result is known to 
change once subleading interactions are taken into account \cite{Bardeen:1989ds}, and hence we take this result as only
indicative that the top-Higgs should have a mass of order 200 - 700 GeV.

\item  The mass splitting between the charged and neutral top-pions is relatively small -- with $\Delta M_\Pi/M_\Pi$
less than about 10\%. We therefore conclude that the minimum amount of isospin violation required
in topcolor (the amount necessary to yield the top-quark mass) need not produce a large mass splitting between
the top-pions.

\item The analysis also confirms that the form of the potential in Eq. (\ref{eq:Vhiggs}), with $\epsilon \simeq 0$, 
correctly summarizes the non-derivative interactions yielding the top-pion and top-Higgs masses and interactions. 
We therefore typically expect $M_{\Pi_t} \lesssim M_{H_t}$, {\it c.f.} Eq. (\ref{eq:pimass}) for small $\sin\omega$.

\end{itemize}

Based on these considerations, in what follows we explore the possibility that the new state at a mass of approximately 125 GeV observed at the LHC is consistent with a neutral pseudoscalar top-pion state.  We consider
 two representative cases: (1) assuming degenerate charged and neutral top-pion masses, $M_{\Pi_t^\pm}=M_{\Pi_t^0}$, 
and  (2) fixing $M_{\Pi_t^0} \approx
125$ GeV and allowing the charged top-pion mass to vary. As discussed above, the first case is that generically expected
in top color models, and the second allows us to illustrate how these results would change if the top-color dynamics includes
additional sources of isospin violation.

\subsection{Scalar Couplings to Fermions}

The couplings of the top-pion and top-Higgs to fermions are model dependent.
Unlike in the standard model, the presence of  two different sources for the quark masses (topcolor and technicolor)
implies that the top-pion and top-Higgs couplings depend on the individual left-handed and right-handed rotations 
in the separate up- and down-quark sectors that relate the {\it topcolor} gauge eigenstates (in which the 
top-pion and top-Higgs couplings are simple) to the mass-eigenstates \cite{Hill:1994hp,Kominis:1995fj,Buchalla:1995dp}. 

For our analysis, we make the following assumptions:

\begin{itemize}

\item Following \cite{Hill:1994hp,Kominis:1995fj,Buchalla:1995dp},
we assume that the top- and bottom-quarks both receive most of their mass as a result of topcolor 
(which would naturally explain why $V_{tb}\simeq 1$), while the other quarks and the leptons receive their masses from the (extended) technicolor sector.
That is, if we were to ``turn off" technicolor electroweak symmetry breaking ($F \to 0$ or $\cos\omega \to 0$)
the top and bottom quarks would have masses close to their observed values, but all other quarks and
the leptons would be massless.

\item The usual CKM angles are related to the difference between the left-handed up- and down-quark rotations which are 
required. Since the observed CKM matrix is non-trivial, it is not possible that {\it both} of the left-handed up- and down-quark 
rotations are trivial. As we show in appendix \ref{sec:flavor}, however, if the observed CKM angles arise predominantly from rotations 
in the left-handed down-quark sector, charged top-pion exchange will lead to unacceptably large contributions to
the process $b \to s \gamma$. We therefore assume that CKM mixing arises from the rotations in the
left-handed up-quark sector.

\item The rotations in the right-handed sector are, {\it a priori}, unconstrained. However, if present, they have the
potential to lead to unacceptably large contributions to  $B^0_d-\bar{B}^0_d$ \cite{Kominis:1995fj} and 
$D^0-\bar{D}^0$ meson mixing. We therefore assume that there is no mixing in the right-handed sector.

\end{itemize}

With these assumptions, to leading order, the flavor-diagonal couplings of the neutral top-pions to the third generation 
fermions\footnote{Couplings to the light quarks and leptons would follow the same pattern as for the $\tau$ lepton, but
will not be needed in what follows.} are
\begin{equation}
\frac{i\,\Pi^0_t}{v} \left[
m_t \cot\omega\, \bar{t}_L t_R + m_b \cot\omega\, \bar{b}_L b_R +
m_\tau \tan\omega\, \bar{\tau}_L \tau_R \right] + h.c.~.
\label{eq:fermioni}
\end{equation}
The mixing in the left-handed up-quark sector will necessarily lead to flavor-changing decays of the neutral top-pion
\cite{He:1998ie,Burdman:1999sr} of the form
\begin{equation}
\frac{i\,\Pi^0_t}{v}\, m_t \cot\omega\, \left[
V^{CKM}_{cb} \bar{c}_L t_R + V^{CKM}_{ub} \bar{u}_L t_R
\right]~.
\label{eq:fermionii}
\end{equation}
The couplings of the top-Higgs to fermions are the scalar analogs of the pseudo-scalar couplings
of the $\Pi^0_t$ listed in Eqs. (\ref{eq:fermioni}) and (\ref{eq:fermionii}) above. 

Similarly, the corresponding charged-pion couplings are of the form\footnote{The coupling of $\Pi^+_t$
to $\bar{t}_R b_L$ gives a potentially large contribution to the process $Z \to b \bar{b}$ \protect\cite{Burdman:1997pf},
which must be compensated for by adjusting the properties of the top-quark \protect\cite{Chivukula:2011ag}.
See the discussion in appendix \protect\ref{sec:flavor}.}
\begin{equation}
\frac{i\sqrt{2} \Pi^+_t}{v} \left[
m_t \cot \omega\, \bar{t}_R b_L + m_b \cot\omega\, \bar{t}_L b_R
+ m_b \cot\omega\, V^{CKM}_{cb} \bar{c}_L b_R
+ m_\tau \tan\omega\, \bar{\nu}_{\tau L} \tau_R +
m_c \tan\omega\, R_{cs} \bar{c}_R s_L\right] + h.c. ~,
\label{eq:fermioniii}
\end{equation}
where $R_{cs}$ is an unknown mixing parameter which, for the purposes of illustration,
we take equal to its maximum value $R_{cs} \simeq \cos\theta_C \simeq 1$.\footnote{If this
coefficient were smaller, this would increase the branching ratio $BR(\Pi^+_t \to \bar{\tau} \nu_\tau)$
which would strengthen the limits in section \protect\ref{sec:chargedpionlimits}.}

The relation between the assumptions made here and the simpler form of the fermion
couplings used in \cite{Chivukula:2009ck} is presented in Appendix \ref{sec:flavor}.

\section{Neutral top-pion phenomenology}
\label{sec:neutralpionlimits}

In this section, we will discuss the phenomenology of the neutral top-pion assuming it has a mass of 125 GeV.   We start by reviewing the couplings and decays, examine the production cross-section, and then discuss various decay modes in light of the LHC data.  More details about the model can be found in Ref.~\cite{Chivukula:2011ag}.


\subsection{Couplings and decays}

The couplings of the neutral top-pion that are most relevant to our analysis are those to $gg,\gamma\gamma, b\bar{b}$, and $\tau\bar{\tau}$.  The couplings to gluon pairs or photon pairs arise from top quark loops (contributions from loops containing heavy top-quark partners would be suppressed by powers of the heavy quark mass).  Those to fermions arise from top color (for $t$ and $b$) and/or extended technicolor dynamics (especially for lighter fermions).  Being a pseudoscalar, the top-pion lacks tree-level couplings to $WW$ and $ZZ$, and the loop induced couplings to these massive gauge bosons are small compared to the dominant ones listed above. These decays do occur through a top-quark loop, and are discussed separately.

\begin{figure}
\resizebox{\textwidth}{!}{
\includegraphics[scale=1]{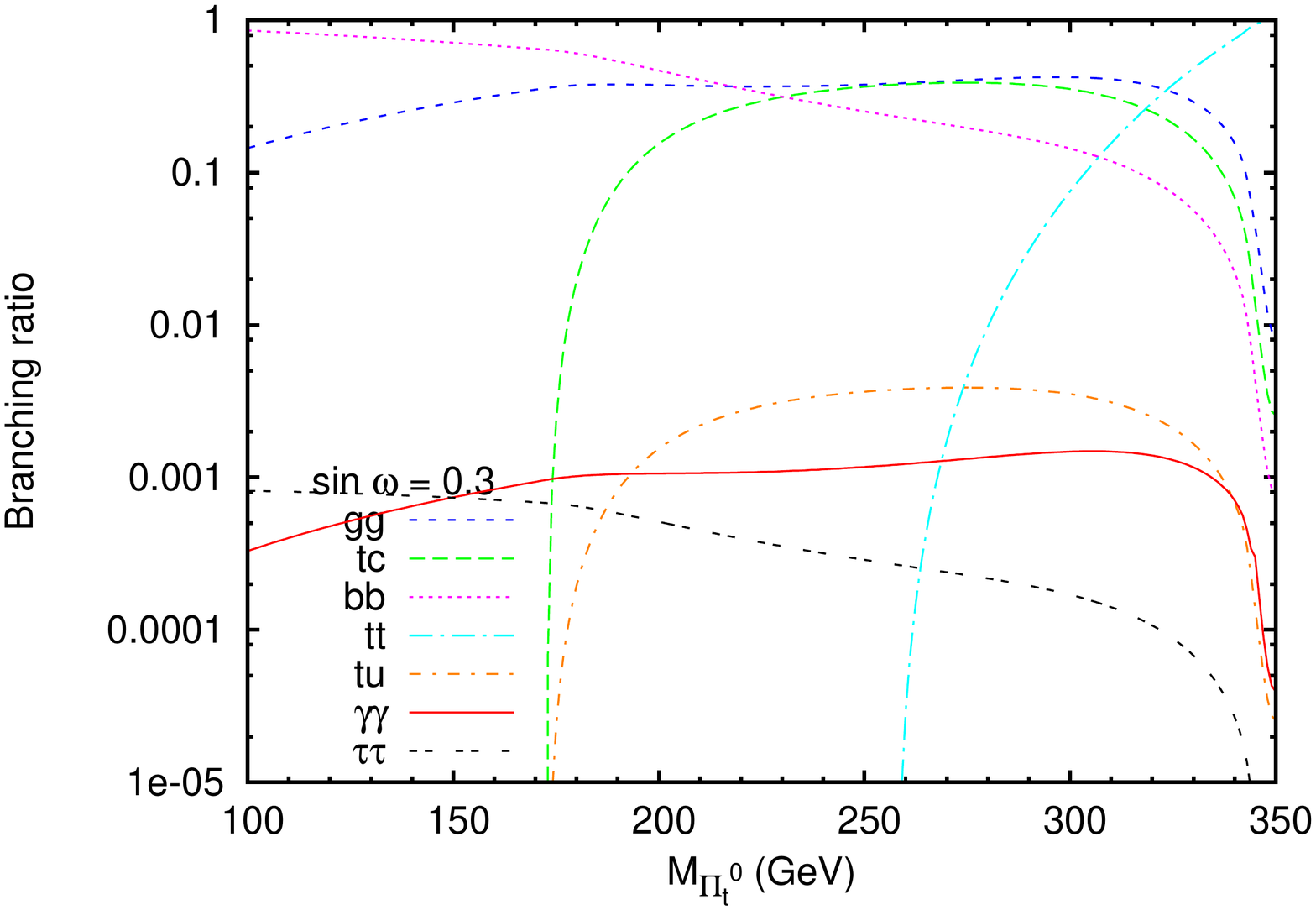}
\includegraphics[scale=1]{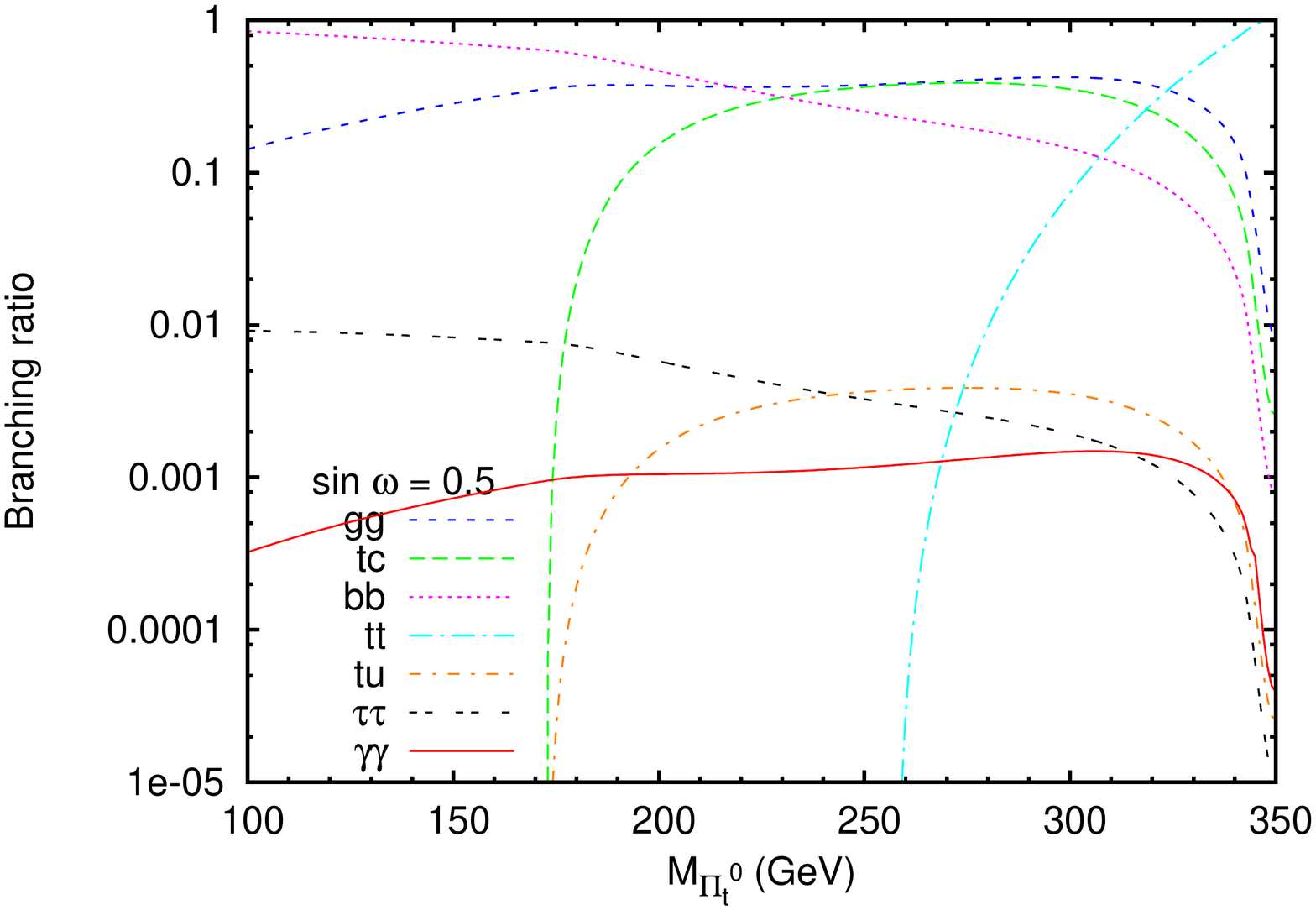}
\includegraphics[scale=1]{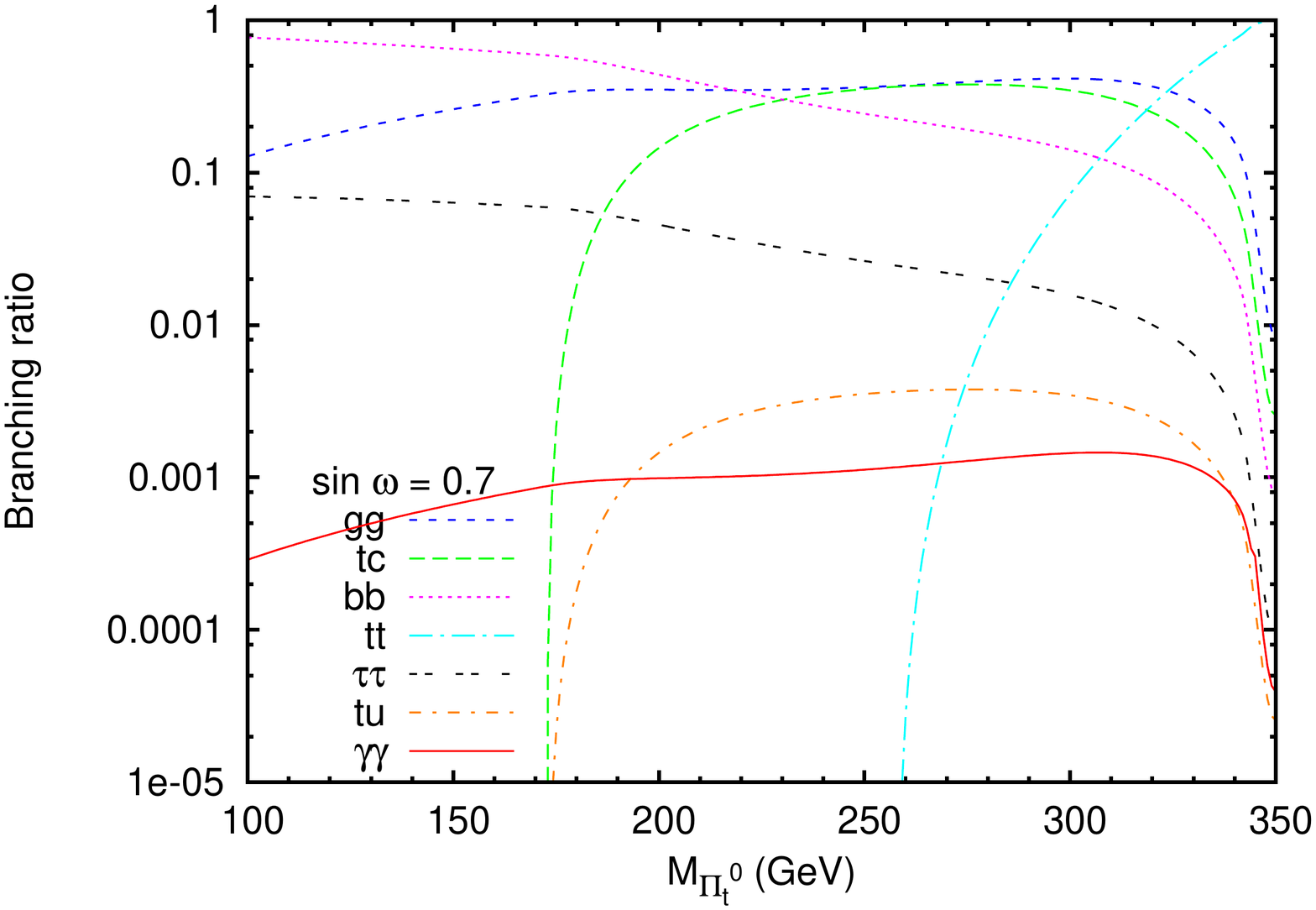}}
\caption{Branching ratios of the $\Pi_t^0$ into its dominant decay modes for $\sin\omega =$ 0.3 (left), 0.5 (center), and 0.7 (right). The order of the curves in the key (from top to bottom) reflects the order of the curves at $M_{\Pi^0_t} = 300$ GeV. 
} 
\label{fig:BRs}
\end{figure}

We have calculated the branching ratios of $\Pi_t^0$ using the MSSM pseudoscalar decay routines in {\tt HDECAY} version 3.531~\cite{HDECAY}, modified to take into account the different fermion coupling structure of Eqs.~(\ref{eq:fermioni})--(\ref{eq:fermionii}) and the absence of superpartners.  The resulting branching ratios are illustrated in Fig.~\ref{fig:BRs} for $\sin\omega = $ 0.3, 0.5 and 0.7.  Decays to $b\bar{b}$ dominate at  low $\Pi^0_t$ mass, with the $gg$ and $tc$ channels becoming important only once $M_{\Pi^0_t} \gtrsim 200$ GeV.  Decays to $t \bar t$ turn on at $M_{\Pi_t} \simeq 2 m_t \simeq 350$~GeV and completely dominate above this mass.  Note that our calculation using {\tt HDECAY} includes decays to off-shell $t \bar t$ below threshold.  As these plots indicate, for a 125 GeV top-pion, only the decay branching ratios to $gg,\gamma\gamma, b\bar{b}$ and $\tau\bar{\tau}$ are significant.

\begin{figure}
\resizebox{\textwidth}{!}{
\includegraphics{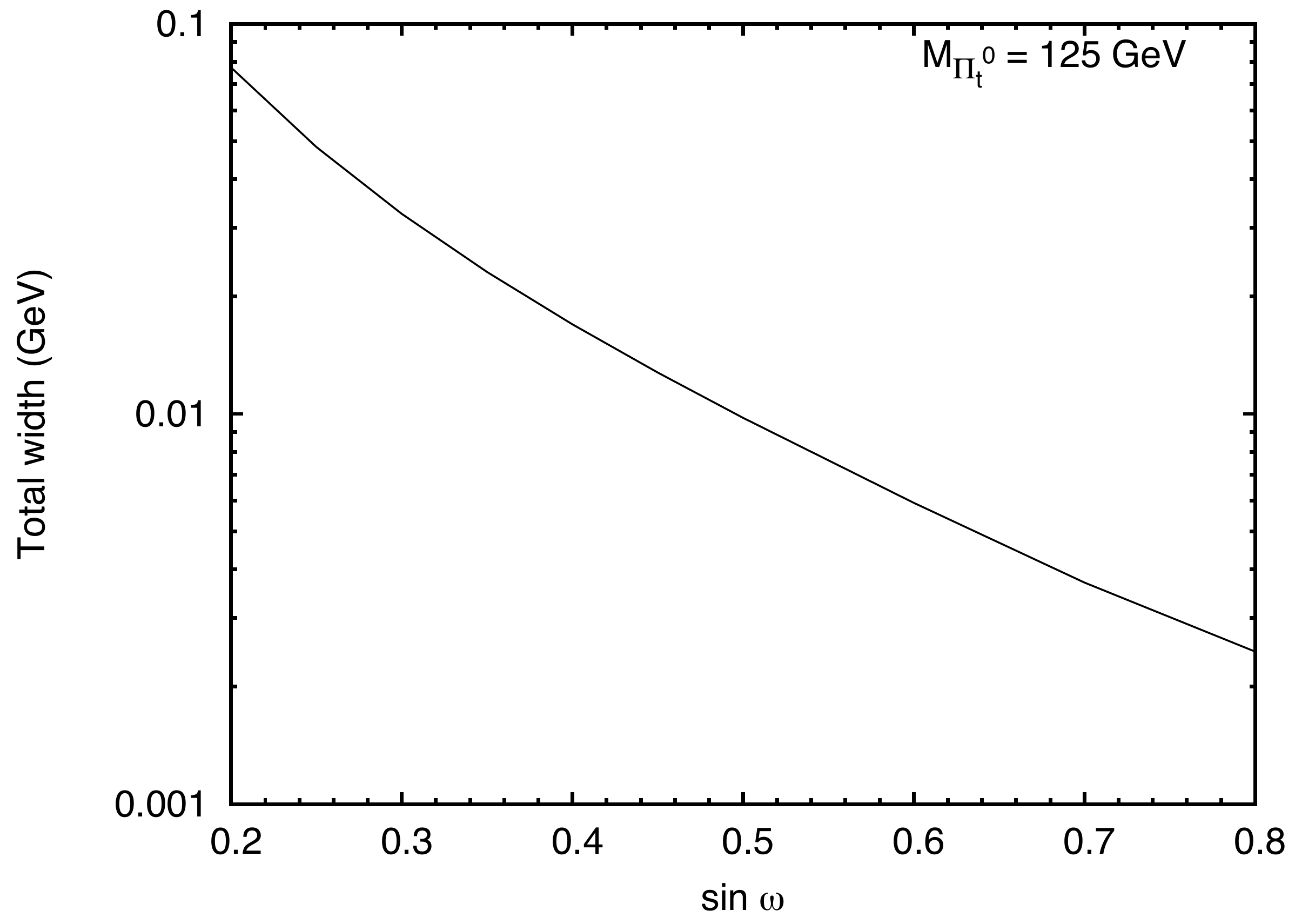}
\includegraphics{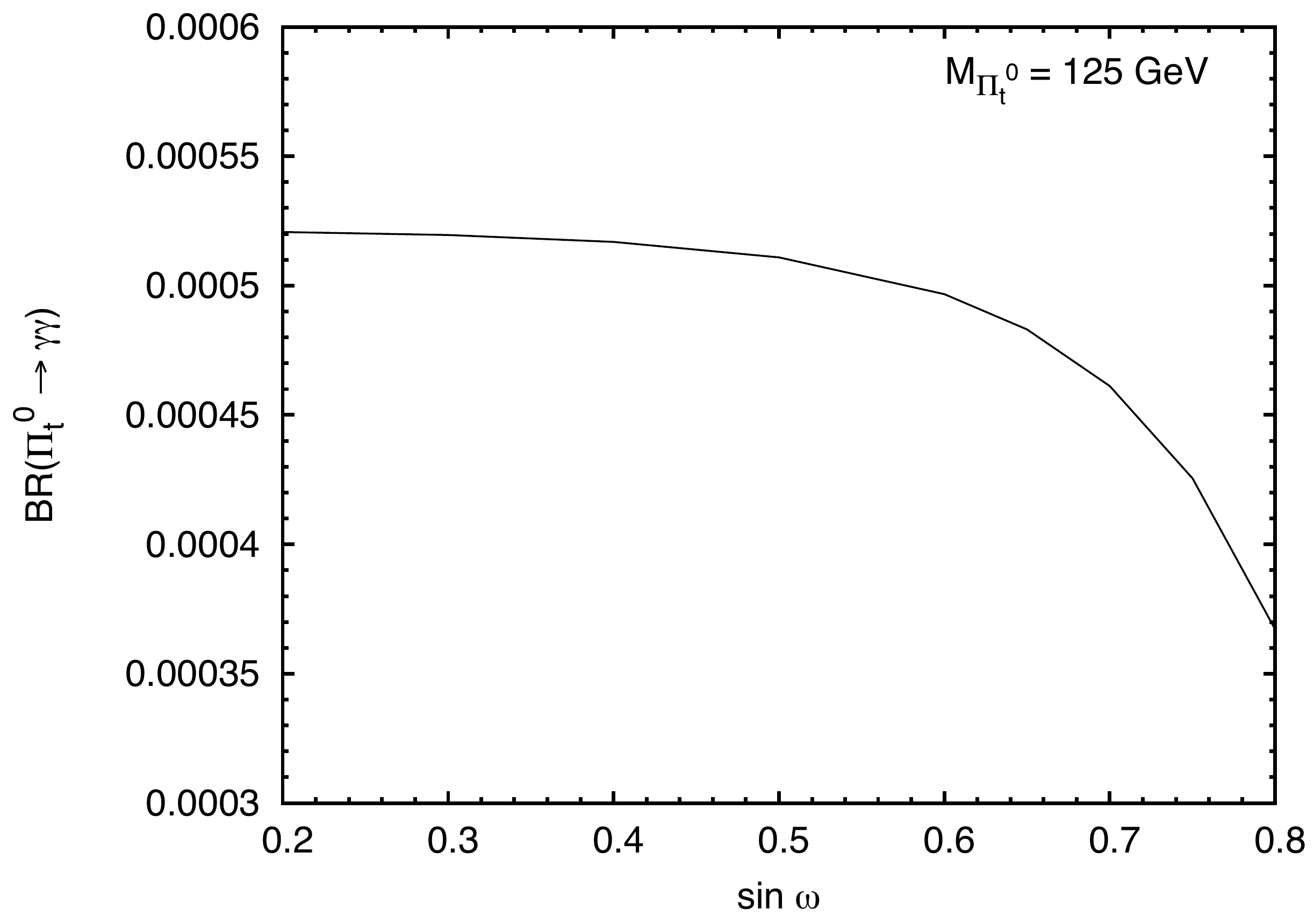}
}
\caption{Total decay width (left) and  branching ratio for $\Pi_t^0 \to \gamma\gamma$ (right) of a 125 GeV $\Pi_t^0$ as a function of $\sin\omega$.  
}
\label{fig:gamtot}
\end{figure}

The total width of $\Pi_t^0$ is shown in the left panel of Fig.~\ref{fig:gamtot} as a function of  $\sin\omega$, with $M_{\Pi_t^0} = 125$ GeV.   Because its mass is well below the $t\bar t$ threshold, the $\Pi_t^0$ remains a narrow resonance with width below 1~GeV for all values of $\sin\omega \geq 0.2$.

In the right panel of Fig.~\ref{fig:gamtot} we display the branching ratio for $\Pi_t^0 \to \gamma\gamma$ as a function of  $\sin\omega$ with $M_{\Pi_t^0} = 125$ GeV.  This branching ratio reaches at most 0.5 parts per mil and is roughly five times smaller than the SM Higgs branching ratio into photons.

\subsection{Production cross-section}

Here, we calculate the production cross-section of the neutral top-pion; in subsequent sub-sections we will compare this prediction to various ATLAS and CMS results to analyze the current and future LHC sensitivity to neutral top-pions.  

The neutral top-pion is produced at the LHC almost exclusively via gluon fusion.  We calculate the cross section for $\Pi_t^0$ production in gluon fusion according to
\begin{equation}
	\sigma(gg \to \Pi_t^0) = \frac{ \left| \sum_f \alpha_f F^A_{1/2}(\tau_f) \right|^2}
	{\left| \sum_f F^H_{1/2}(\tau_f) \right|^2}
	\times \sigma(gg \to H_{\rm SM}),
	\label{eq:xsec}
\end{equation}
where in the sum over fermions we include\footnote{Technifermion loops do not contribute to top-pion production because the $SU(2)_{weak} \times [SU(3)]^2$ anomaly vanishes for any realistic technicolor theory. } $t$, $b$ and $c$; also $\alpha_t = \alpha_b = \cot\omega$ and $\alpha_c = \tan\omega$.  Here the fermion loop functions $F^H_{1/2}(\tau)$ and $F^A_{1/2}(\tau)$, for scalars and pseudoscalars respectively, are given by~\cite{Gunion:1989we}:
\begin{eqnarray}
	F^H_{1/2}&=&-2\tau\left[1+(1-\tau)f(\tau) \right], \nonumber \\
	F^A_{1/2}&=&-2\tau f(\tau),
	\label{eq:F}
\end{eqnarray}
where $\tau_f=4 m_f^2/M_{\Pi}^2$ and
\begin{equation}
f(\tau) =
  \begin{cases}
   \left[\sin^{-1}\left(\sqrt{1/\tau} \right) \right]^2 & \text{if } \tau \geq 1 \\
   -\frac{1}{4}\left[\textrm{ln}(\eta_+/\eta_-)-i\pi  \right]^2      & \text{if } \tau < 1,
  \end{cases}
\label{eqn:tau}
\end{equation}
with $\eta_{\pm}=(1\pm\sqrt{1-\tau})$.  In the limit of a heavy fermion in the loop, $F^H_{1/2} \to -4/3$ and $F^A_{1/2} \to -2$.

We take the SM gluon-fusion Higgs production cross section $\sigma(gg \to H_{\rm SM})$ from Ref.~\cite{xsecwg} for the 7~TeV LHC.  This SM Higgs cross section includes the state-of-the-art radiative corrections, which boost the cross section by a substantial factor $\sim 2$.  Our cross section in Eq.~(\ref{eq:xsec}) relies on the equality of the $k$-factors for pseudoscalar production and scalar production.  In fact, because most of the QCD $k$-factor comes from real radiation, this equality has been shown to hold to within 20\%, as illustrated in \cite{Spira:1995rr}.

\subsection{Current and Prospective Limits from the Diphoton channel}

\begin{figure}[h]
\includegraphics[scale=0.3]{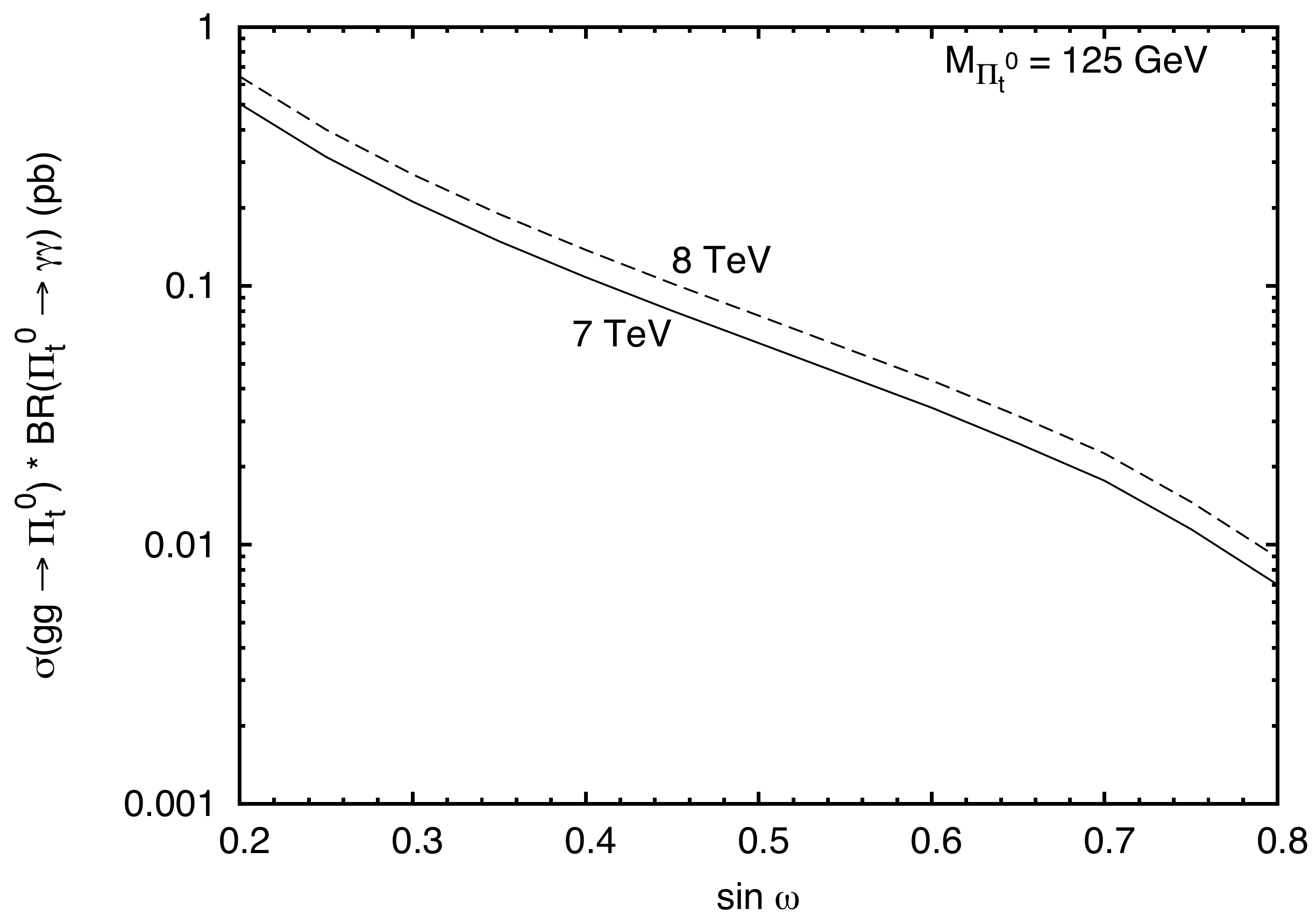}
\includegraphics[scale=0.3]{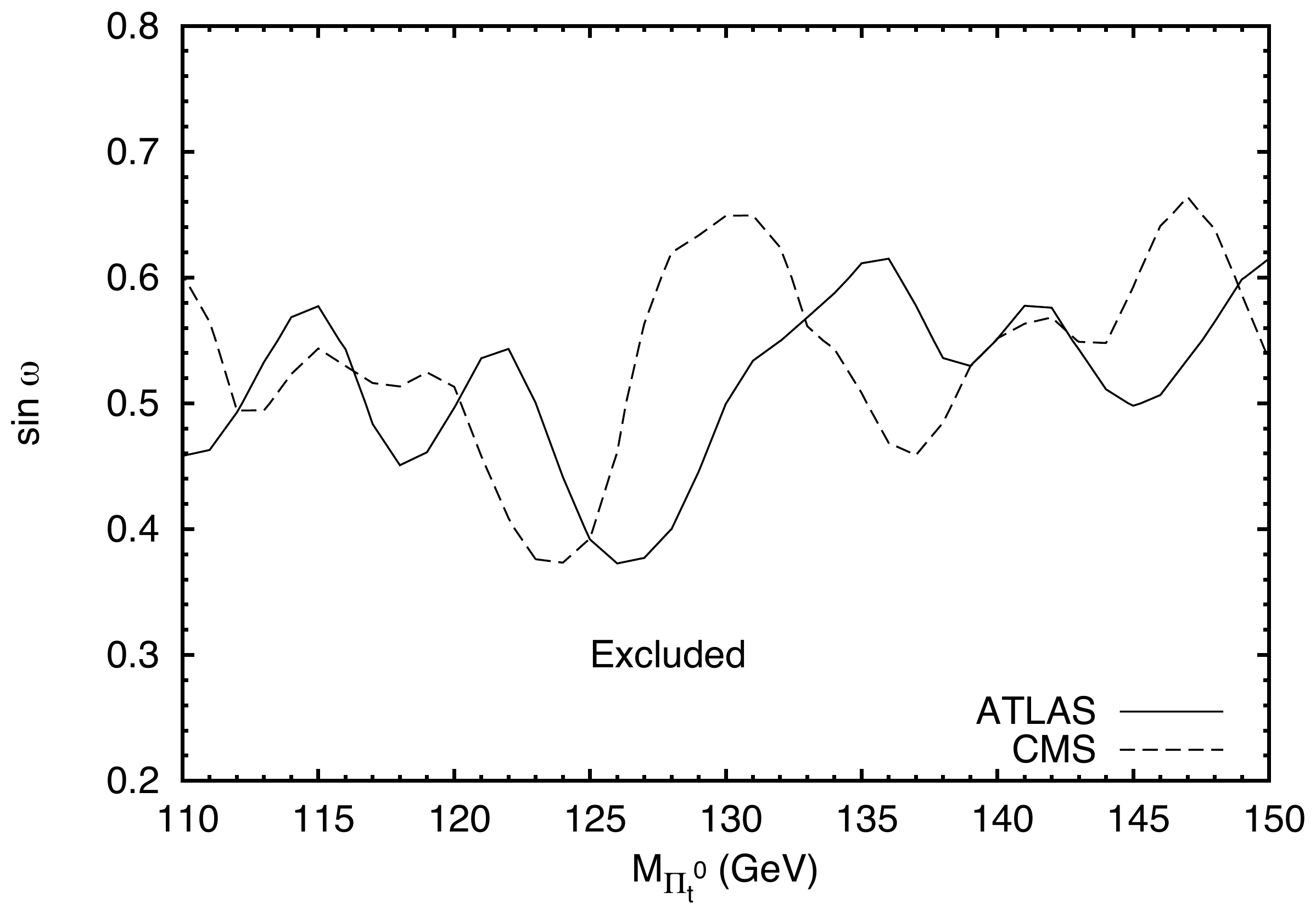}
\caption{Left:  Cross section times branching ratio for $gg \to \Pi_t^0 \to \gamma\gamma$ for a 125 GeV top-pion at the
7 and 8 TeV LHC. Right: 95\% confidence level exclusion limits in the $M_{\Pi^0}$ vs. $\sin\omega$ plane, extrapolated from the LHC SM Higgs search limits in the $\gamma\gamma$ channel with 4.8--4.9~fb$^{-1}$ at 7~TeV from Refs.~\cite{atlas:2012sk,Chatrchyan:2012tw}.    
}
\label{fig:pigagareal}
\end{figure}

The diphoton decay channel has played a leading role in LHC searches for the Standard Model Higgs boson.  Although this is not the dominant decay mode for the neutral top-pion, it would certainly be highly visible in the LHC detectors.   We have calculated  $\sigma(gg \to \Pi_t^0) \times {\rm BR}(\Pi_t^0 \to \gamma\gamma)$ and our results are shown as a function of $\sin\omega$ (fixing $M_{\Pi^0_t} = 125$ GeV) in the left-hand pane of Fig.~\ref{fig:pigagareal}.  The signal rate is largest for small $\sin\omega$, due to the enhancement of both the $\Pi_t^0$ production cross section and the branching ratio to $\gamma\gamma$ at small $\sin\omega$.  

The LHC SM Higgs searches in Refs.~\cite{atlas:2012sk,Chatrchyan:2012tw} have exclusion sensitivity to $\gamma\gamma$ resonances with a cross section of order 50~fb for resonance masses between 110 and 150~GeV.  We find that this excludes a neutral top-pion in this mass range with $\sin\omega \lesssim 0.4-0.5$.  We show the excluded region in the right plot in Fig.~\ref{fig:brXbr}, based on the 95\% confidence level limit on $\sigma/\sigma_{\rm SM}$ in the $\gamma\gamma$ channel alone for the SM Higgs from Refs.~\cite{atlas:2012sk,Chatrchyan:2012tw}; those limits are based on 4.9~fb$^{-1}$ (ATLAS) and 4.8~fb$^{-1}$ (CMS) at 7~TeV.
We translated the LHC results into bounds on our model by comparing the CMS and ATLAS limits on $\sigma/\sigma_{\rm SM}$ with
\begin{equation}
	\frac{\sigma}{\sigma_{\rm SM}} =
	\frac{\sigma(gg \to \Pi_t^0)\times {\rm BR}(\Pi_t^0 \to \gamma\gamma)}
	{[\sigma(gg \to H_{\rm SM}) + \sigma({\rm VBF} \to H_{\rm SM}) ]
	\times {\rm BR}(H_{\rm SM} \to \gamma\gamma)},
\end{equation}
where $\sigma(gg \to \Pi_t^0)$ is obtained from Eq.~(\ref{eq:xsec}) and $\sigma({\rm VBF} \to H_{\rm SM})$ is the SM Higgs production cross section via vector boson fusion (VBF).  Note that the CMS analysis includes a contribution from a dedicated VBF search topology channel, which would not be present for the top-pion.  The ATLAS analysis does not include a dedicated VBF channel and is thus more directly applicable to the top-pion.  However, the inclusion of the dedicated VBF search channel by CMS does not appear to significantly affect our results: the limits are consistent with each other in excluding low values of $\sin\omega$.

Both CMS and ATLAS observe a new state  with a mass of about 125~GeV decaying to diphotons whose properties appear to be consistent with those of a SM Higgs boson.  However, the observed diphoton rate is nearly twice that expected for a SM Higgs~\cite{CMSgaga,ATLASgaga}, which also makes the excess consistent with a neutral top-pion with $\sin\omega \simeq 0.5$, as shown in the right-hand pane of Fig.~\ref{fig:pigagareal}.

\subsection{Decays to $ZZ$, $Z\gamma$ and $WW$}

It is interesting to consider how one would be able to distinguish a neutral top-pion from a SM Higgs boson once more data is in hand.   The SM Higgs has tree-level couplings to $W^+W^-$ and $ZZ$, while couplings to $\gamma\gamma$ arise only at one loop.  In contrast, being a pseudoscalar, $\Pi^0_t$ does not have tree-level couplings to $W^+W^-$ or $ZZ$ \cite{Frandsen:2012rj,Low:2012rj,Coleppa:2012eh}.  It can, however, have couplings to $W^+W^-$, $ZZ$, and $Z\gamma$ at one loop.  Ref.~\cite{Coleppa:2012eh} considered the possibility that the loop-induced pseudoscalar coupling to the $SU(2)$ and hypercharge gauge bosons can account for the observed $\gamma\gamma$ and $4\ell$ signal. Essentially, the strategy consisted in adjusting the relative value of the $SU(2)$ and hypercharge gauge couplings so that the equation
\begin{equation}
	\frac{\Gamma^c(H \to ZZ^* \to 4e)}{\Gamma(H \to \gamma\gamma)}
	\ = \
	\frac{\Gamma^c(\phi \to 4e)}{\Gamma(\phi \to \gamma\gamma)}.
	\label{eq:replacement}
\end{equation}
is satisfied. Here, $\phi$ refers to the pseudoscalar and the superscript $c$ means the quantities are computed with the experimental cuts imposed. It was shown that the $Z\gamma^{*}$ contribution to the $4\ell$ signal completely dominates the $ZZ^{*}$ contribution, in direct contrast to the SM case where the $Z\gamma^{*}$ contribution is negligible. Fixing the ratio of the coupling strengths this way leads to a well-defined prediction \cite{Coleppa:2012eh}:
\begin{equation}
R_{Z\gamma/\gamma\gamma}^{\phi} 
	\equiv \frac{\Gamma(\phi \to Z\gamma)}{\Gamma(\phi \to \gamma\gamma)}
	=  121.
\label{eq:ratio}
\end{equation}

Thus, in order to see if the top-pion can generate the experimentally required signal strength, it suffices to compute the ratio of the partial widths to $Z\gamma$ and $\gamma\gamma$ and compare with the number in Eq.~(\ref{eq:ratio}). This number turns out to be $\approx$ 0.02 for the top-pion.\footnote{We note that this is independent of $\sin\omega$, which cancels out in the ratio. $\sin\omega$ in our model is analogous to the parameter $c$ in \cite{Coleppa:2012eh} - one that can be tuned to adjust the production cross-section to the proper value.} Thus, we conclude that the top-pion cannot generate the observed ratio of the $4\ell$ to $\gamma\gamma$ rates. Though this might seem to be a problem for models with strong top-quark dynamics in general, we point out that the $4\ell$ signal involves very few events and conclusions about the viability of our model based on this observation should be postponed until higher-statistics results are available from the ATLAS and CMS collaborations.

\subsection{Limits from the Ditau Channel}

Searching for light neutral top-pions decaying to $\tau\tau$ is difficult because of the large Drell-Yan background. For scalars that are produced in part by vector boson fusion, the sensitivity can be enhanced by implementing cuts that preferentially select the VBF channel, but unfortunately this option is not available for pseudoscalars like the top-pion.

Looking specifically at the case where the neutral top-pion is responsible for the diphoton excess at 125~GeV ($M_{\Pi_t^0} = 125$~GeV and $\sin\omega \simeq 0.5$, corresponding to an enhancement in the $\gamma\gamma$ channel by about a factor of 2 compared to the SM Higgs prediction), then we expect a $\tau\tau$ signal rate, from the gluon fusion channel, approximately equal to that of the SM Higgs.  This is about a factor of 3 below the sensitivity of the $H_{\rm SM} \to \tau\tau$ search from ATLAS~\cite{ATLASHtautau} (4.7~fb$^{-1}$ at 7~TeV).  However, that ATLAS analysis includes events in a ``Higgs plus two jet" event category, corresponding to VBF production, to improve the sensitivity to the SM Higgs, so this limit does not directly apply to the neutral pseudoscalar top-pion of our model.   The papers \cite{CMSgaga,CMS:2012gu,ATLASgaga,ATLAS:2012gk} reporting the discovery of a new scalar in the diphoton channel do analyze data from the ditau channel, but neither finds conclusive evidence that the new state decays to tau lepton pairs.  In the future, perhaps a dedicated search focused on the gluon fusion production channel would be sensitive to the $\Pi^0_t$.

\section{Charged top-pion phenomenology}
\label{sec:chargedpionlimits}

Charged top-pions would be pair-produced via electroweak processes at LHC and their dominant decay channels are hadronic.  Therefore a direct search for $\Pi^+_t$ would be hampered by a combination of low cross-section and high backgrounds.  The main constraints on these states presently come from top quark decays.

\subsection{Branching ratios}

\begin{figure}
\resizebox{\textwidth}{!}{
\includegraphics{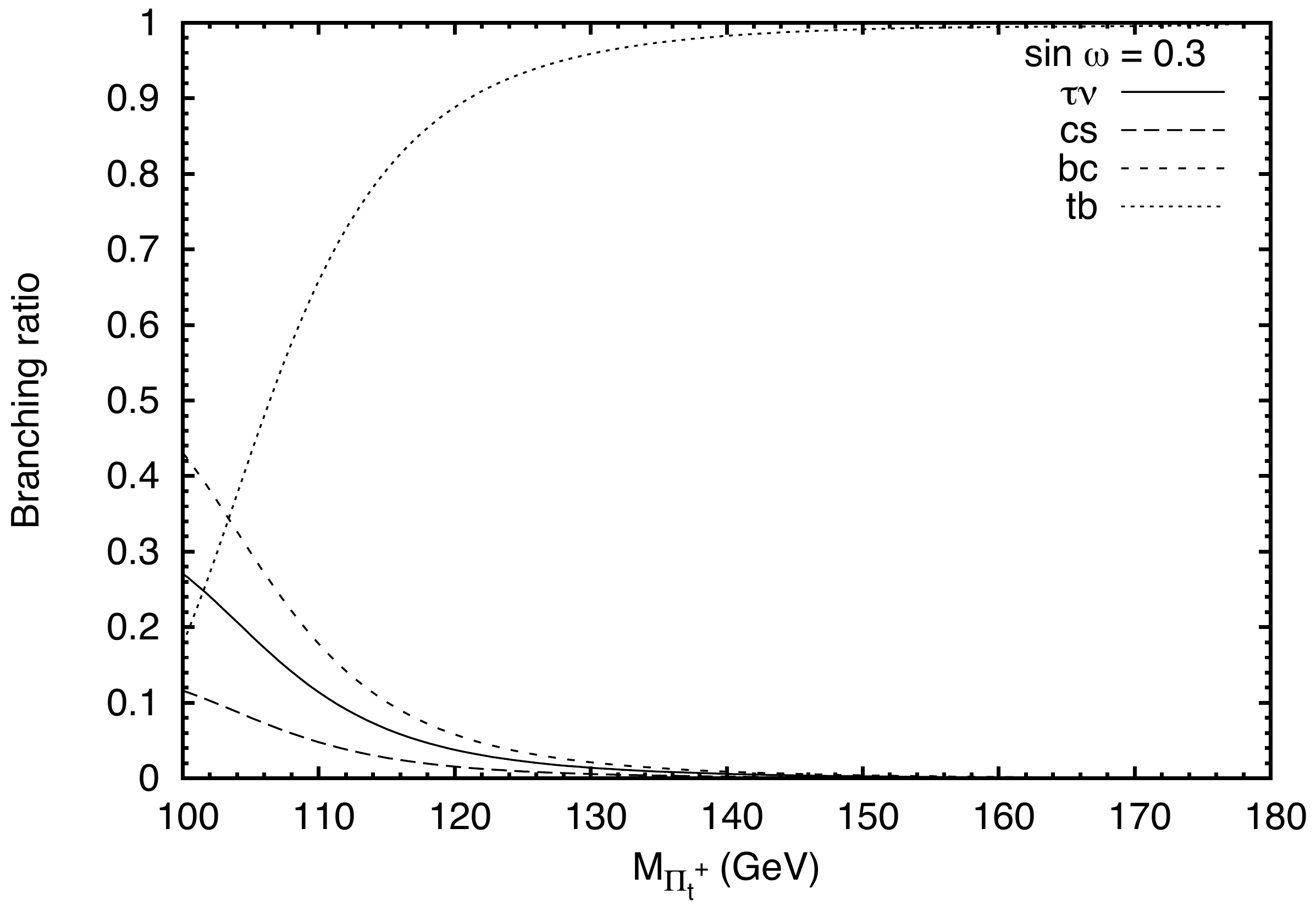}
\includegraphics{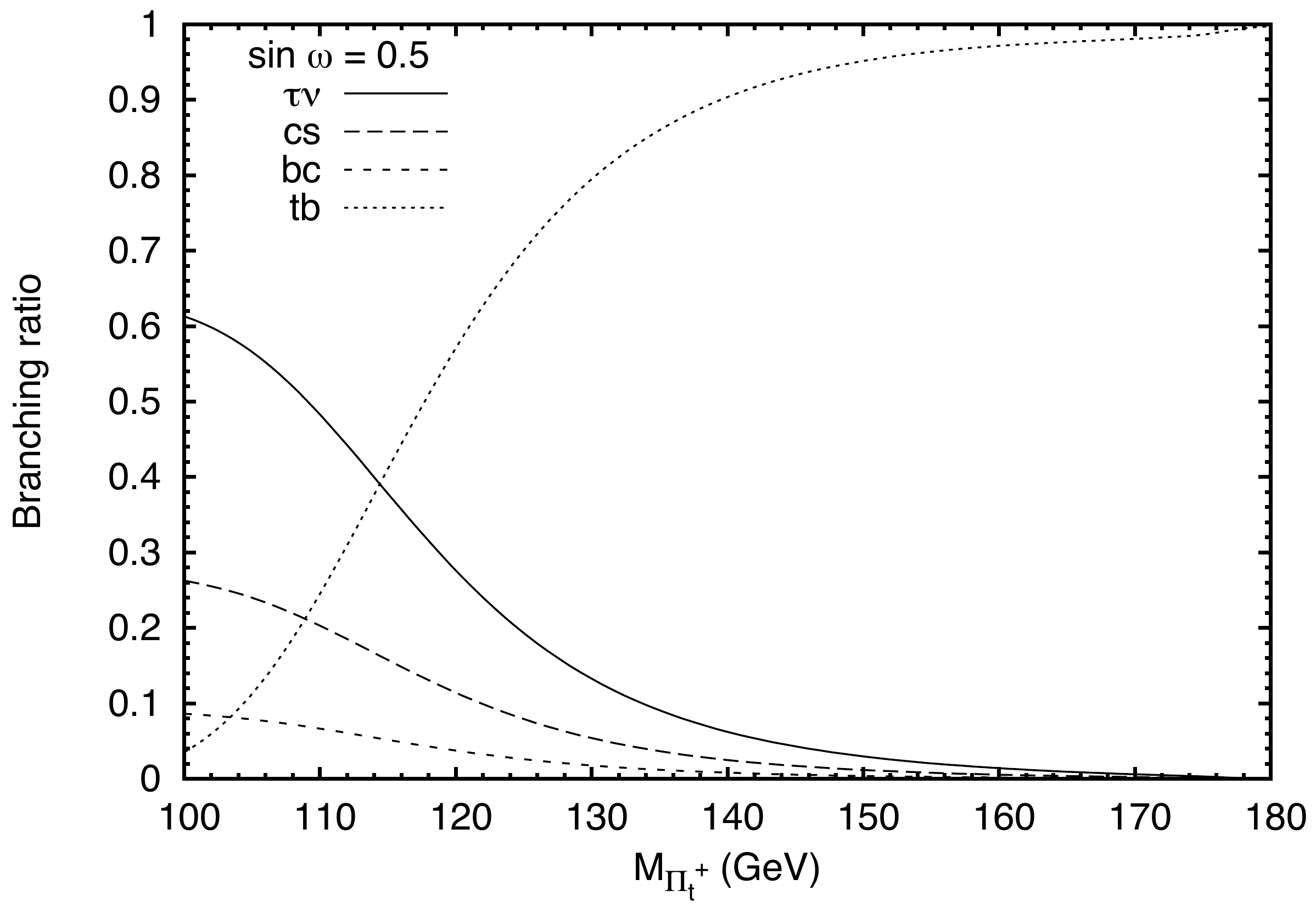}
\includegraphics{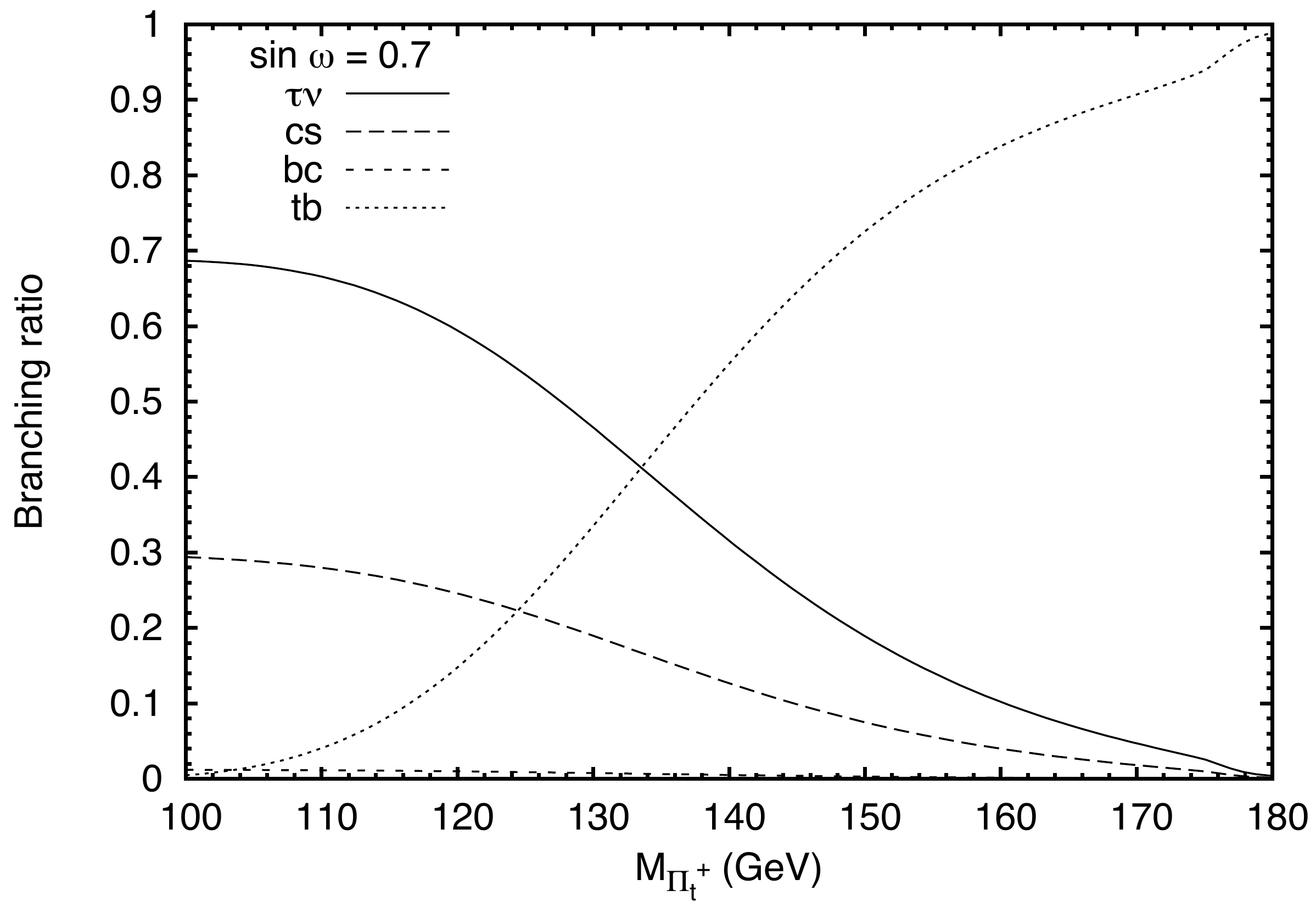}}
\caption{\label{fig:pi+brs} Branching ratios of the charged top-pion to the dominant final states, for $\sin\omega = 0.3$, 0.5, and 0.7 (left to right).  We include off-shell decays to $t^*b$ and also off-shell decays to $\Pi^0_t W^+$ assuming $M_{\Pi^0_t} = 125$ GeV.  These branching ratios were computing using a modified version of {\tt HDECAY}~\cite{HDECAY}.}
\end{figure}

We plot the branching ratios of the charged top-pion as a function of $M_{\Pi^\pm_t}$ in Fig.~\ref{fig:pi+brs}, assuming that the mass of the neutral top-pion is fixed at 125 GeV. For top-pion masses below $m_t$, the dominant decays are into $\tau \nu$ and offshell $t^*b$; their relative rates depend on the top-pion mass and $\sin\omega$.  The decay to $cs$ has a branching ratio a little less than half that of $\tau\nu$; the rate for $bc$ is many times smaller when $\sin\omega \gtrsim 0.5$.  For masses above $m_t$, decays to $tb$ overwhelmingly dominate.  

If $M_{\Pi^+_t} > M_{\Pi^0_t}$ then the off-shell decay $\Pi^+_t \to \Pi^0_t W^{+*}$ becomes possible.   As this branching ratio never exceeds 5\%, it is phenomenologically unimportant for our purposes.  

Since the charged top-pion seldom decays to the neutral top-pion even when kinematically allowed to do so, Fig.~\ref{fig:pi+brs} also gives a good sense of the branching ratios of the charged top-pion for the case in which the top-pions are degenerate.

\subsection{Limits from $t \to \Pi^+ b$}

The ATLAS collaboration has searched for evidence of charged scalars in top quark decays using  4.6~fb$^{-1}$ of data gathered at 7~TeV \cite{ATLASH+}.  Because this was motivated as a search for the charged Higgs of the MSSM, which decays almost exclusively to $\tau\nu$ for large $\tan\beta$, their search assumed that the charged scalar would decay only to $\tau\nu$.  Specifically, they set a limit on $B \equiv {\rm BR}(t \to H^+ b)$, assuming that ${\rm BR}(H^+ \to \tau \nu) = 1$.  The latter assumption is built directly into their analysis in that they scale the simulated cross section for SM $t \bar t$ background, in which $t \to Wb$, by $(1 - B)^2$.

\begin{figure}
\resizebox{\textwidth}{!}{
\includegraphics{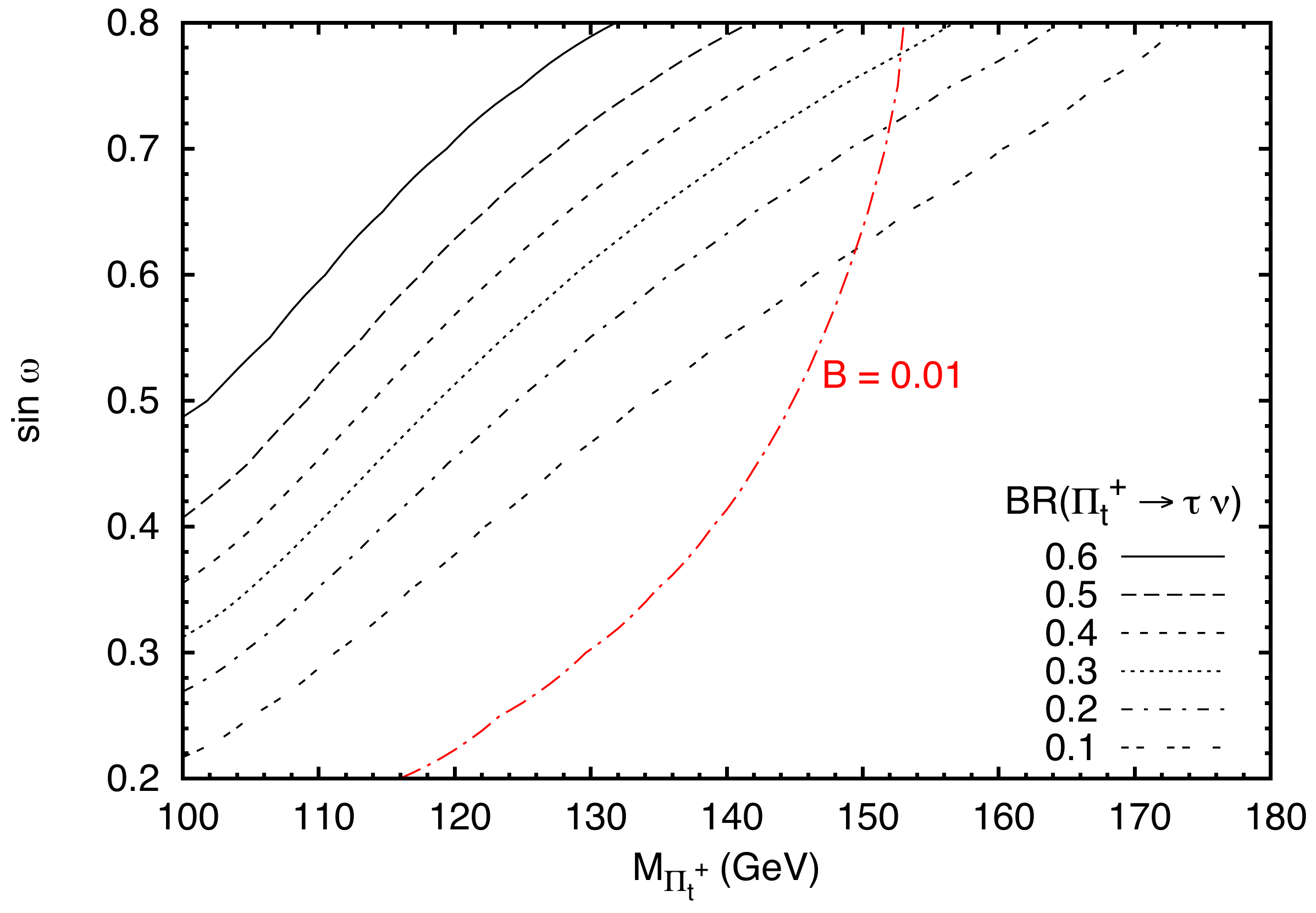}
\includegraphics{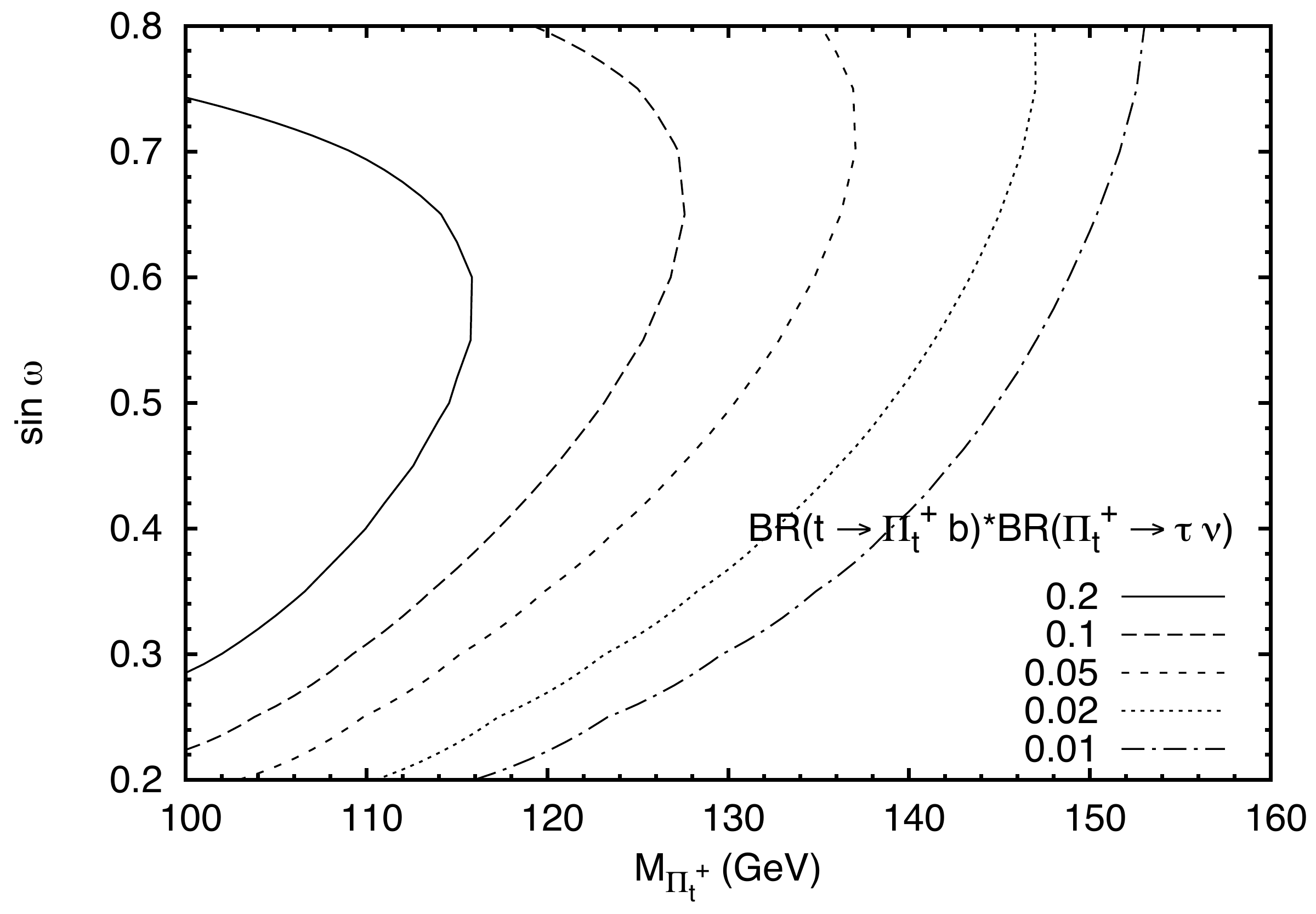}}
\caption{\label{fig:brtaunu} Left:  Contours of ${\rm BR}(\Pi^+_t \to \tau\nu)$ as a function of $M_{\Pi^+_t}$ and $\sin\omega$.  Right: 
Contours of ${\rm BR}(t \to \Pi^+_t b) \times {\rm BR}(\Pi^+_t \to \tau \nu)$ as a function of $M_{\Pi^+_t}$ and $\sin\omega$.  We interpret the ATLAS $t \to H^+ b$ search~\cite{ATLASH+} to exclude $B \equiv {\rm BR}(t \to \Pi^+_t b) \times {\rm BR}(\Pi^+_t \to \tau \nu) > 0.01$. }
\end{figure}

The conclusions of the ATLAS $t \to H^+ b$ analysis cannot be directly applied to the charged top-pion because ${\rm BR}(\Pi^+_t \to \tau \nu) \neq 1$, as can be seen in Fig.~\ref{fig:pi+brs}.  In fact, as also illustrated in the left pane of Fig.~\ref{fig:brtaunu}, the value of ${\rm BR}(\Pi^+_t \to \tau\nu)$ ranges from a maximum of about 0.7 for a relatively light $\Pi^+_t$ and large $\sin\omega$, to close to zero for a heavier top-pion and lower $\sin\omega$ (due to the competing $t^*b$ decay).  

Nevertheless, we can adapt the ATLAS $t \to H^+b$ limits to extract information about the charged top-pion.  The charged top-pion signal is the same as that for the charged Higgs studied in Ref.~\cite{ATLASH+}, provided that the parameter $B$ is replaced by ${\rm BR}(t \to \Pi^+_t b) \times {\rm BR}(\Pi^+_t \to \tau\nu)$.   We calculated the top quark decay branching ratio at tree level neglecting the bottom quark mass, using
\begin{equation}
	{\rm BR}(t \to \Pi^+_t b) = \frac{\cot^2\omega (1 - M_{\Pi^+}^2/m_t^2)^2}
	{(1 + 2 M_W^2/m_t^2)(1 - M_W^2/m_t^2)^2 + \cot^2\omega (1 - M_{\Pi^+}^2/m_t^2)^2}.
\end{equation}
We have calculated the $\Pi^+_t$ decay branching ratios using a modified version of {\tt HDECAY}~\cite{HDECAY} as discussed before.   Combining these branching fractions, we show contours of ${\rm BR}(t \to \Pi^+_t b) \times {\rm BR}(\Pi^+_t \to \tau \nu)$ in the $M_{\Pi^+_t}$ vs. $\sin\omega$ plane in the right-hand pane of Fig.~\ref{fig:brtaunu}.

\begin{figure}
\resizebox{\textwidth}{!}{
\includegraphics{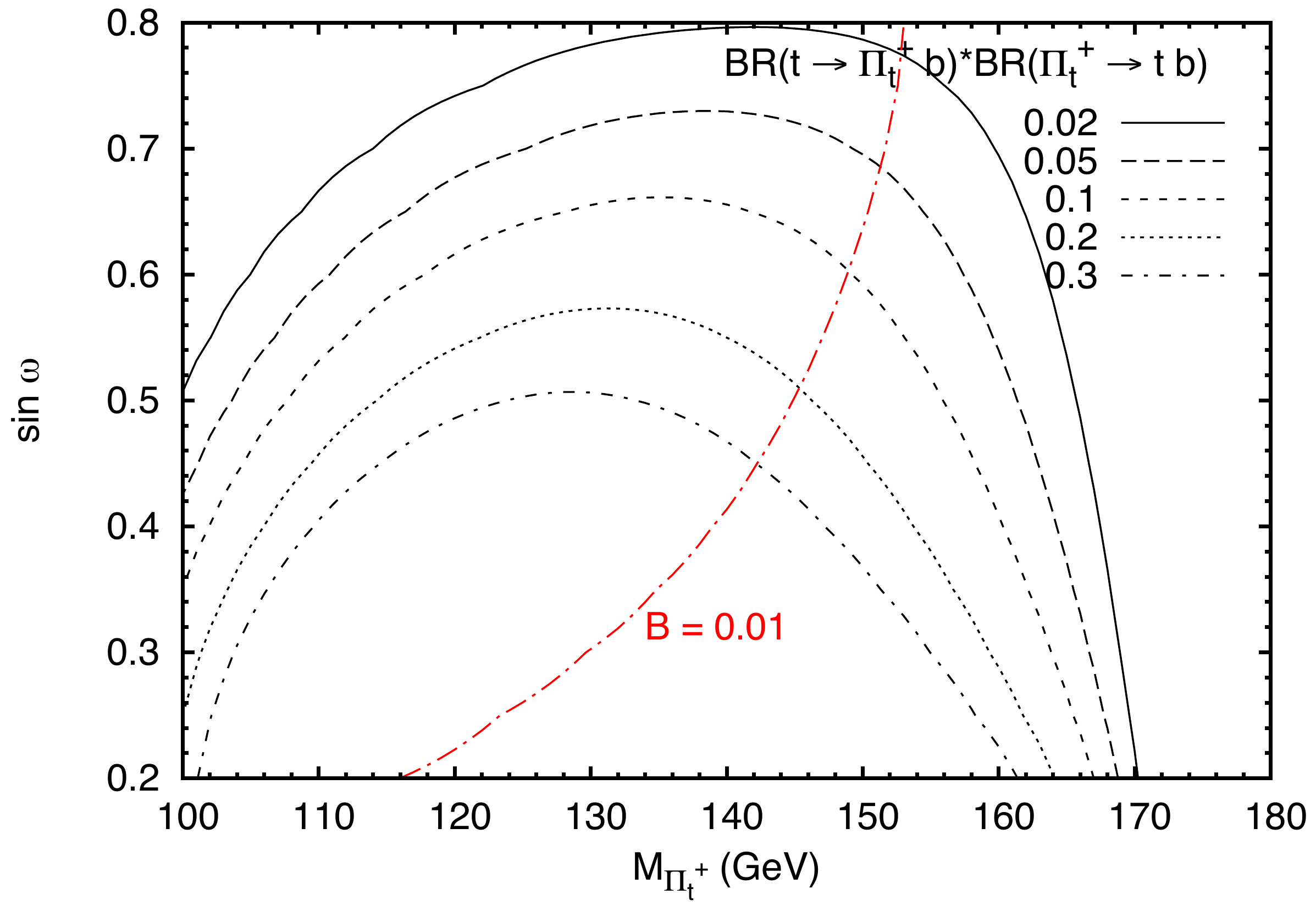}
\includegraphics{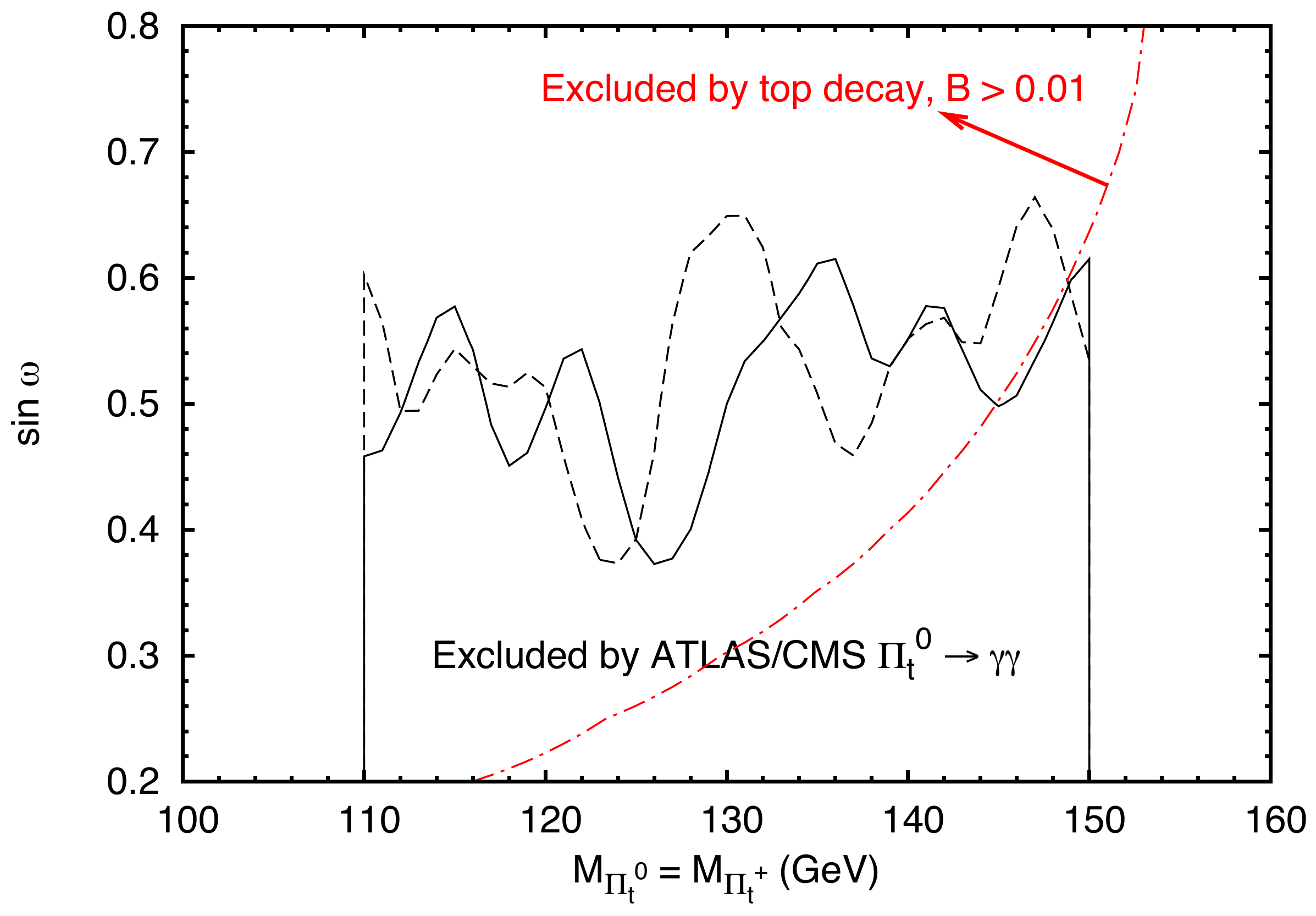}}
\caption{\label{fig:brXbr} Left: 
Contours of ${\rm BR}(t \to \Pi^+_t b) \times {\rm BR}(\Pi^+ \to t^* b)$ as a function of $M_{\Pi^+_t}$ and $\sin\omega$. On each plot, only the region of parameter space to the right of the red dot-dashed curve labeled ``B=0.01" is still allowed by data on top-quark decays, as shown in Fig.~\protect\ref{fig:brtaunu}.
 Right: comparison of the ATLAS $t \to H^+ b$ exclusion and the ATLAS (solid) and CMS (dashed) $\Pi^0_t \to \gamma\gamma$ limits \protect\cite{atlas:2012sk,Chatrchyan:2012tw}, assuming degenerate $\Pi^0_t$ and $\Pi^+_t$.}
\end{figure}

However, the ``SM-like" top-pair events to which the signal events are compared in setting a limit on exotic top decays will no longer include only $t\to Wb$ events.  This sample will now potentially contain events in which a top-pion decays to $bt^*$, yielding $t \to \Pi^+ b \to W^+ b \bar b b$, where the $W^+b$ comes from the off-shell top quark.  While the kinematic features of these top decays will differ from those of SM decays, the events may be similar enough to be picked up in the SM top quark sample.  To see how common these events are, we show contours of ${\rm BR}(t \to \Pi^+_t b) \times {\rm BR}(\Pi^+ \to t^* b)$ in the plane of $M_{\Pi^+_t}$ and $\sin\omega$ in the left-hand panel of Fig.~\ref{fig:brXbr}.  The product of branching ratios can be significant: it lies above 0.3 for $\sin\omega < 0.45$ and $M_{\Pi^+_t} \sim 140$~GeV.  In this case more than half of all $t \bar t$ events would contain at least one top quark decaying to $\Pi^{\pm}_t b$ followed by $\Pi^{\pm}_t \to t^*b \to W^{\pm} b \bar b$; we suspect that this could distort kinematic distributions and $b$-tag rates in the $t \bar t$ sample enough to be noticed. Similarly, for $\sin\omega = 0.5$ and $M_{\Pi^+_t} \simeq 145$~GeV, we find ${\rm BR}(t \to \Pi^+_t b) \times {\rm BR}(\Pi^+ \to t^* b) \simeq 0.2$, leading to about 40\% of $t \bar t$ events containing at least one top quark decaying to $\Pi^{\pm}_t b$ followed by $\Pi^{\pm}_t \to t^*b \to W^{\pm} b \bar b$.

While deliberately distinguishing these $t \to \Pi^+ b \to W^+ b \bar b b$ events from SM top quark decays would require a dedicated analysis, in the meantime, we can make the conservative assumption that all of these events will be included in the ``SM-like" sample.  When this is the case, the comparison between exotic and SM-like events gives a conservative upper limit on ${\rm BR}(t \to \Pi^+_t b) \times {\rm BR}(\Pi^+_t \to \tau \nu)$.  When some of these events are not picked up in the SM-like sample, the true upper bound on the product of branching fractions is actually even stronger. 

We are now ready to determine the constraints on our model.  Reference~\cite{ATLASH+} sets an upper bound on $B \equiv {\rm BR}(t \to H^+ b)$ (with ${\rm BR}(H^+ \to \tau \nu) = 1$) of $B \lesssim 0.05$ for $M_{H^+} = 90$~GeV, falling to $B \lesssim 0.01$ for $M_{H^+} = 120$--160~GeV.  Therefore we can take the right-most contour in the left-hand pane of Fig.~\ref{fig:brXbr} as the rough exclusion limit on $\Pi^+_t$ from this search channel.  This excludes charged top-pion masses below about 118, 140, 149, and 153~GeV for $\sin\omega = 0.2$, 0.4, 0.6, and 0.8, respectively.   We have overlaid this exclusion curve on the plots in Fig.~\ref{fig:brtaunu} to make it easier to see what values of ${\rm BR}(\Pi^+_t \to \tau\nu)$ and ${\rm BR}(t \to \Pi^+_t b) \times {\rm BR}(\Pi^+ \to t^* b)$ are still allowed in our model.  Note, for instance,  that, for $\sin\omega \lesssim 0.6$, the region of parameter space where the $B \leq 0.01$ limit falls has ${\rm BR}(\Pi_t^+ \to \tau\nu) < 0.1$.  

Finally, examining the right-hand pane of Fig.~\ref{fig:brXbr},  we see that if the new state observed in diphotons at around 125~GeV is to be interpreted as a $\Pi^0_t$ (with the event rate yielding $\sin\omega = 0.5$), then we would interpret the combination of the diphoton data from Refs.~\cite{atlas:2012sk,Chatrchyan:2012tw} and ATLAS search~\cite{ATLASH+} for $t \to H^+ b$ with $H^+ \to \tau\nu$ as jointly constraining the $\Pi^+_t$ to be heavier than about 145~GeV.   

Therefore, the only phenomenologically viable case involves non-degenerate top-pions.  As discussed in
detail in Appendix~\ref{appendix:NJL}, however, this differs from the standard expectation in top color models and implies that new sources of isospin violation would have to be present.

\section{Top-Higgs phenomenology}
\label{sec:tophiggslimits}

In addition to the top-pion states discussed above, models in which the top-quark plays a direct role in electroweak
symmetry breaking contain a ``top-Higgs" state. Such a state is expected to have a mass greater
than about 200 GeV, and we have
previously demonstrated  \cite{Chivukula:2011dg} that
such a top-Higgs state would produce $ZZ$ and $WW$ signals much {\it larger} than those characteristic of a SM Higgs of the same mass when decays to pairs of top-pions are kinematically forbidden. In this
section we consider the constraints on the top-Higgs state assuming that the neutral top-pion is the new
boson discovered at the LHC.

The couplings of the top-Higgs, along with its decay widths to the most relevant channels $WW$, $ZZ$, $t\bar{t}$, $\Pi_t^{\pm}W^{\mp}$, $\Pi_t^0Z$, $\Pi_t^{+}\Pi_t^{-}$, and $\Pi_t^0 \Pi_t^0$, are given in detail in Ref.~\cite{Chivukula:2011dg}.  For completeness, we reproduce the formulas for the key decay widths in Appendix~\ref{sec:tophiggsdecays}, along with the ratio between the LHC production cross-sections for the top-Higgs and the SM Higgs. We will first establish the current mass limits on the top-Higgs based on data from the ATLAS and CMS experiments. We then comment on the discovery prospects for the top-Higgs in the channel $H_t \to \Pi^0_t Z$ at the 14 TeV LHC.

Reference~\cite{Chivukula:2011dg} used the combined SM Higgs limits from the LHC to determine the excluded range of top-Higgs masses as a function of $\sin\omega$, for various values of the top-pion mass.  In the mass range of interest, the LHC limits come entirely from the SM Higgs decays into $WW$ and $ZZ$, and so are directly applicable to the top-Higgs after rescaling by the appropriate ratios of production cross section and decay branching ratios.  The limits of Ref.~\cite{Chivukula:2011dg} used ATLAS results with 1.0--2.3~fb$^{-1}$ and CMS results with 1.1--1.7~fb$^{-1}$ of integrated luminosity at 7~TeV.  Here we update the limits using the more recent CMS SM Higgs search results based on 4.6--4.8~fb$^{-1}$ at 7~TeV and also consider how the limits translate to the case where the charged and neutral top-pions are not degenerate.

\begin{figure}[!t]
\resizebox{0.55\textwidth}{!}{\includegraphics{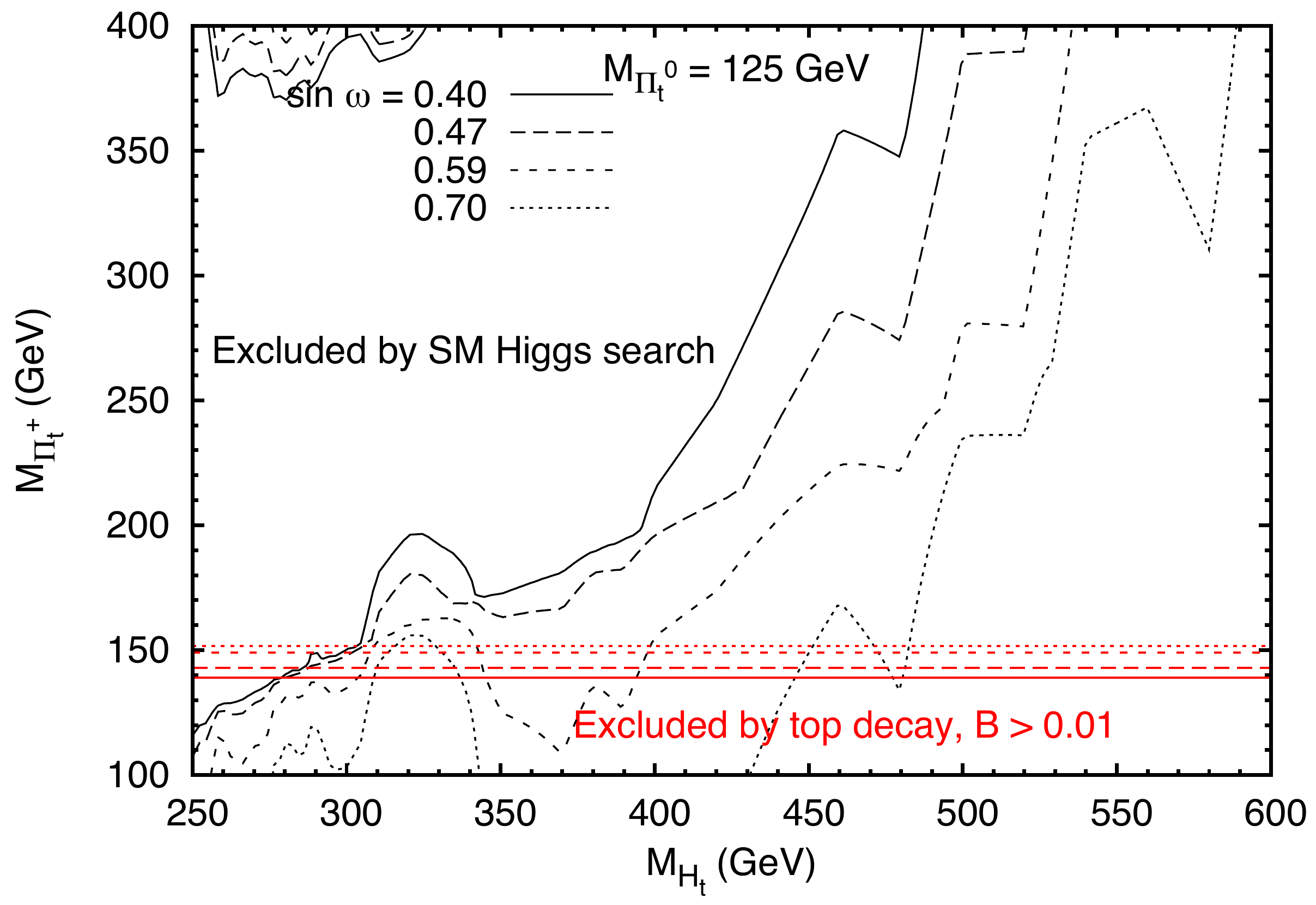}}
\caption{The CMS exclusion contours for $M_{H_t}$ from searches for the SM Higgs in $WW$ and $ZZ$ final states~\cite{Chatrchyan:2012tx}, as a function of $M_{\Pi^+_t}$ for the special case $M_{\Pi^0_t} = 125$~GeV.  We show $\sin\omega$ values of 0.40 (solid lines), 0.47 (long-dashed lines), 0.59 (short-dashed lines), and 0.70 (dotted lines), which correspond to a rate for the 125~GeV $\Pi^0_t$ in the $\gamma\gamma$ channel relative to that of the SM Higgs of $\sigma/\sigma_{\rm SM} = 3.0$, 2.0, 1.0, and 0.5, respectively.  
The horizontal (red) lines show our lower bound on $M_{\Pi^+_t}$ from the ATLAS $t \to H^+ b$ search~\cite{ATLASH+} for the same four $\sin\omega$ values.}
\label{fig:Ht125}
\end{figure}

Fig.~\ref{fig:Ht125} shows how the top-Higgs exclusion curves behave for a variety of $\sin\omega$ values. Given that light charged top-pions are excluded by  the ATLAS search \cite{ATLASH+} for $t \to H^+ b$, the top-Higgs cannot have a mass lower than about 250-300 GeV.  

There is also a theoretical bound to bear in mind. For small values of $\sin\omega$, the top-Higgs couplings violate perturbativity for sufficiently high $H_t$ masses, when the decay channels to two tops and two top-pions open up; roughly speaking this occurs when the top-Higgs width exceeds its mass. For $\sin\omega \approx 0.4$ we find this constrains the top-Higgs mass to lie below about 600 GeV, while for $\sin\omega \geq 0.5$, perturbativity considerations do not constrain the region of interest.

\begin{figure}
\includegraphics[scale=1.3]{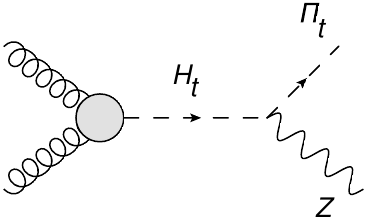}
\caption{The production of a neutral top-pion and a $Z$ from an $s$-channel top-Higgs.}
\label{fig:hpz-feynman}
\end{figure}

Many of the decay channels that are available to a heavy top-Higgs result in hadronic final states with large SM backgrounds.  A potential exception is $H_t \to Z \Pi^0_t \to \ell\ell\gamma\gamma$ as shown in Fig.~\ref{fig:hpz-feynman}.  Assuming the state discovered at 125 GeV is the neutral top-pion, one can then take advantage of the $H_{t}\Pi_{t}^0Z$ coupling, and look for the top-Higgs in the process $pp\rightarrow Z\Pi_t^0\rightarrow \ell\ell\gamma\gamma$ by using an invariant mass cut on the diphotons to cull background.   We find that discovery in the allowed parameter space (see Fig.~\ref{fig:Ht125}) is not possible for $\sin\omega$ values of 0.7 and above in this channel.  Even for lower values of $\sin\omega$, the minimum integrated luminosity required for a $5\sigma$ discovery at the 14 TeV LHC in this mode is 100 $fb^{-1}$ and a luminosity several times greater is required in most of the $M_{H_t}$ vs $M_{\Pi^+_t}$ plane.  Therefore, we conclude that this will not be a realistic discovery mode for the top-Higgs in the case of a light neutral top-pion. The most promising search channels for the top-Higgs therefore remain the $WW$ and $ZZ$ final states as used in the SM Higgs search.

\section{Conclusions }
\label{sec:conclusion}

In this paper we have analyzed the phenomenology of the top-pion and top-Higgs states in models with strong top dynamics, and have translated the present LHC constraints on the SM Higgs into bounds on these scalar states.  

We have seen that it is possible for the observed excess in the $H_{SM} \to \gamma\gamma$ search channel to correspond to a neutral top-pion of mass $M_{\Pi^0_t}=125$ GeV.  Based on the size of the cross-section observed \cite{CMSgaga,CMS:2012gu,ATLASgaga,ATLAS:2012gk}, the corresponding value of $\sin\omega$ would be approximately 0.5.   Because $\Pi^0_t$ is a pseudo scalar, however, models of strong top dynamics do not predict a visible signal in the $ZZ \to 4 \ell$ channel or the $WW$ channel, nor a diphoton signal in the vector boson fusion production channel, nor any associated production of the  125~GeV object with a $W$ or $Z$.     Therefore, as additional data is accumulated, we would expect the diphoton resonance to continue to grow in significance, the initial signals in the $ZZ \to 4\ell$ and $WW$ channels to fade away, and the dijet-tagged diphoton signal to persist only at a level consistent with dijet-tagged $gg \to \Pi^0_t$ rather than dijet-tagged vector boson fusion events.   Moreover, in the context of these models, we would also expect that a signal in the ditau decay channel would be present but less visible for the $\Pi^0_t$ than for the SM Higgs.

For the range of model parameters where the neutral top-pion can account for the LHC diphoton signal, searches for non-standard top-quark decays to charged scalar plus bottom quark exclude charged top-pions with masses up to about 145 GeV (as in the left-hand panel of Fig. \ref{fig:brXbr}).  These searches continue to become more sensitive as the decay properties
of the top-quark are measured more accurately.  As a result, if the neutral top-pion has a mass of 125 GeV, it cannot
be degenerate with the charged top-pion, as one would more typically expect in models of strong top dynamics.  Instead,  the model must contain substantial isospin violation to produce this top-pion mass splitting.   

We have also updated limits on the top-Higgs.  Our results show that current LHC searches for the SM Higgs in $WW$ and $ZZ$ exclude the existence of a top-Higgs state up to masses of order 300 GeV, with some dependence on the charged top-pion mass and
$\sin\omega$ as shown in Fig. \ref{fig:Ht125}.  

The implication is that current searches at the LHC strongly constrain theories with strong top-dynamics.  The top triangle moose model interpolates \cite{Chivukula:2011dg} between a variety of strong top dynamics models as the value of $\sin\omega$ varies between about 0.2 and 0.8, the range studied in this paper.  In the context of strong top-dynamics, the new boson observed at the LHC is too light to be the top-Higgs \cite{Chivukula:2011dg}. Instead, the diphoton signal can be produced by a neutral top-pion of the appropriate mass and couplings, assuming that one constructs a theory including additional isospin violation, but in this case we would not expect a significant signal in the $ZZ \to 4\ell$ channel.  This last stipulation is problematic since both LHC experiments report a 3$\sigma$ signal in the four-lepton channel with the current data set.  Moreover, if the diphoton signal corresponds to a neutral top-pion, then the theoretical context cannot be the most familiar part of the top-triangle-moose parameter space in which $0.2 \leq \sin\omega \leq 0.5$, the top-pions are degenerate, and the top-Higgs has a mass of order $2 m_t$: i.e. the portion of the parameter space corresponding to classic TC2 models.  Rather, the context would be the less-explored region in which $\sin\omega$ is of order 0.5 or greater, the top-pions have a substantial mass splitting, and the top-Higgs is heavier: i.e. a model in which the strong top dynamics are of the top seesaw form.      

We anticipate that additional LHC data will provide further clarity about the nature of the diphoton resonance and its possible connection to strong top dynamics.

\begin{acknowledgments}
B.C.\ and H.E.L.\ were supported by the Natural Sciences and Engineering Research Council of Canada.  R.S.C.\ and E.H.S.\ were supported, in part, by the U.S.\ National Science Foundation under Grant No.\ PHY-0854889 and acknowledge the hospitality of the Aspen Center for Physics where part of this work was completed.  P.I. was supported by the Thailand Development and Promotion of Science and Technology Talents Project (DPST).  A.M.\ was supported by Fermilab operated by Fermi Research Alliance, LLC, under Contract No.\ DE-AC02-07CH11359 with the U.S.\ Department of Energy.  J.R. was supported by the China Scholarship Council.
\end{acknowledgments}

\appendix

\section{TC2 in the NJL Approximation}
\label{appendix:NJL}

In this appendix, we calculate the top-Higgs and top-pion spectrum
in topcolor assisted technicolor (TC2) models \cite{Hill:1994hp}, using the Nambu--Jona-Lasinio (NJL)
\cite{Nambu:1961tp} approximation for the topcolor dynamics. On phenomenological grounds \cite{Braam:2007pm},
we expect the  ``cutoff" $\Lambda$ of the NJL topcolor theory (which is of order the mass of the
gauge-bosons of the topcolor model, {\it i.e.} the top-gluon and $Z'$) to be much higher
than the technicolor scale $\Lambda_{TC}$, which is  of order 1 TeV.  We can therefore construct the 
low-energy theory which we use to compute the scalar spectrum in two stages. 

First, as described in the next section, we integrate out the strong topcolor-induced four-fermion 
operators using the Nambu--Jona-Lasinio approximation, and construct an effective theory involving
a composite top-Higgs field coupled to the third-generation quarks and the technifermions.
This effective theory will be valid at energies below the topcolor cutoff and above the scale
at which the technicolor interactions become strong. Next, as described in the third section,
we match to an effective technicolor chiral Lagrangian valid at low energies. 
 In the fourth section we use this effective Lagrangian to compute the scalar spectrum of the theory.
Custodial isospin violation is necessarily present in the theory so as to explain the top-bottom mass 
difference. In the fifth section we consider what constraints the limits on the custodial isospin violating
parameter $\Delta T$ place on the parameters of the model, and what these restrictions imply for the
scalar mass spectrum. In the last section, we consider the mass splitting between the charged- and 
neutral-top-pions.

\subsection{TC2 Dynamics}

In the NJL approximation,\footnote{The NJL approximation \protect\cite{Nambu:1961tp,Bardeen:1989ds,Chivukula:1998uf} involves two
parts. First we approximate the effects of exchange of heavy top-color gauge bosons by four-fermion contact interactions
and include only those parts of the interaction responsible for coupling left- and right-handed fermion currents.
Second, as discussed below, we analyze the effect of these interactions in the ``fermion bubble" approximation. Here, and in the following,
we also neglect additional TC2 interactions involving the right-handed bottom quark.}
the interactions of interest in this model include
\begin{equation}\label{eq:TC2ETC4fermion}
\frac{g_t^2}{\Lambda^2}(\bar\psi_{L0}t_{R2})(\bar{t}_{R2}\psi_{L0})+\frac{\eta{g_t^2}}{\Lambda^2}\left[(\bar\psi_{L0}t_{R2})(\bar{U}_{R}Q_L)+h.c.\right]
\end{equation}
where the first four-fermion operator is the traditional topcolor interaction responsible for top-quark condensation and the
second, arising from ETC interactions \cite{Eichten:1979ah,Dimopoulos:1979es}, couples the
top-quark to the weak-doublet and singlet technifermions \cite{Weinberg:1979bn,Susskind:1978ms}
$Q_L$ and $U_R$. Here $g_t$ and $\Lambda$
represent the top-color coupling and cutoff, respectively. We expect the second operator to arise from ETC interactions at a scale
larger than $\Lambda$, and for convenience we characterize the strength of these interactions (relative to topcolor) through the small dimensionless parameter $\eta$. All weak, color, and technicolor indices implicit in Eq. (\ref{eq:TC2ETC4fermion}) are summed.

For strong $g_t$, we expect that the topcolor interactions will give rise to a bound electroweak scalar
state with the quantum numbers of the standard model Higgs boson. In the NJL approximation, this
may be seen directly. The interactions Eq. (\ref{eq:TC2ETC4fermion}) may be recast as
\begin{equation}\label{eq:TC2rewritten}
\frac{g_t^2}{\Lambda^2}\left[\bar\psi_{L0}t_{R2}+\eta\bar{Q}_{L}U_{R}\right]\left[\bar{t}_{R2}\psi_{L0}+\eta\bar{U}_{R}Q_L\right]-\frac{\eta^2g_t^2}{\Lambda^2}(\bar{Q}_{L}U_{R})(\bar{U}_{R}Q_L)~,
\end{equation}
which, following \cite{Bardeen:1989ds}, may be rewritten in terms of an auxiliary electroweak doublet scalar
field $\Phi$ (with $SU(2) \times U(1)$ quantum numbers $2_{-1/2}$)
\begin{equation}\label{eq:TC2ETCYukawa}
-\Lambda^2\Phi^{\dag}\Phi-g_t\left[(\bar\psi_{L0}t_{R2}+\eta\bar{Q}_{L}U_{R})\Phi+h.c.\right]
-\frac{\eta^2g_t^2}{\Lambda^2}(\bar{Q}_{L}U_{R})(\bar{U}_{R}Q_L)~.
\end{equation}

\begin{figure}[th]
  \centering
    \includegraphics[width=5cm]{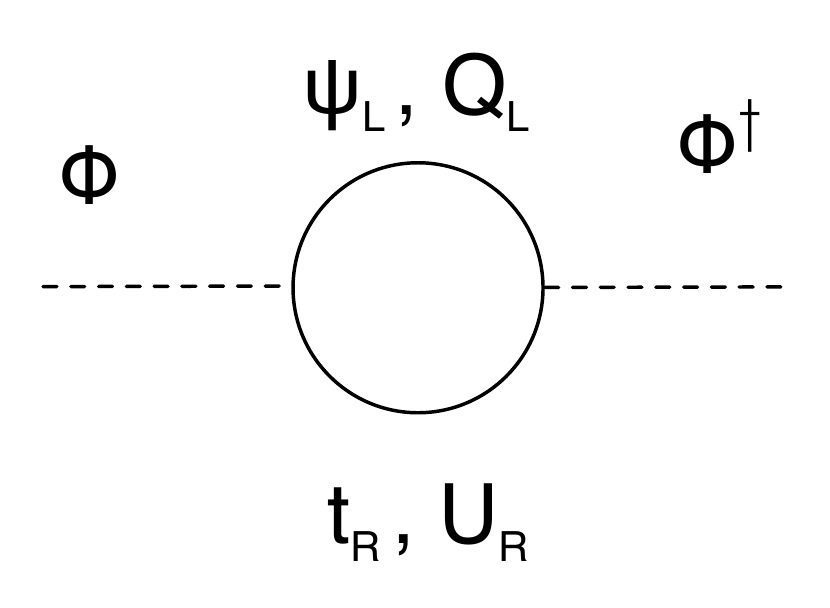}
    \hskip30pt
    \includegraphics[width=6cm]{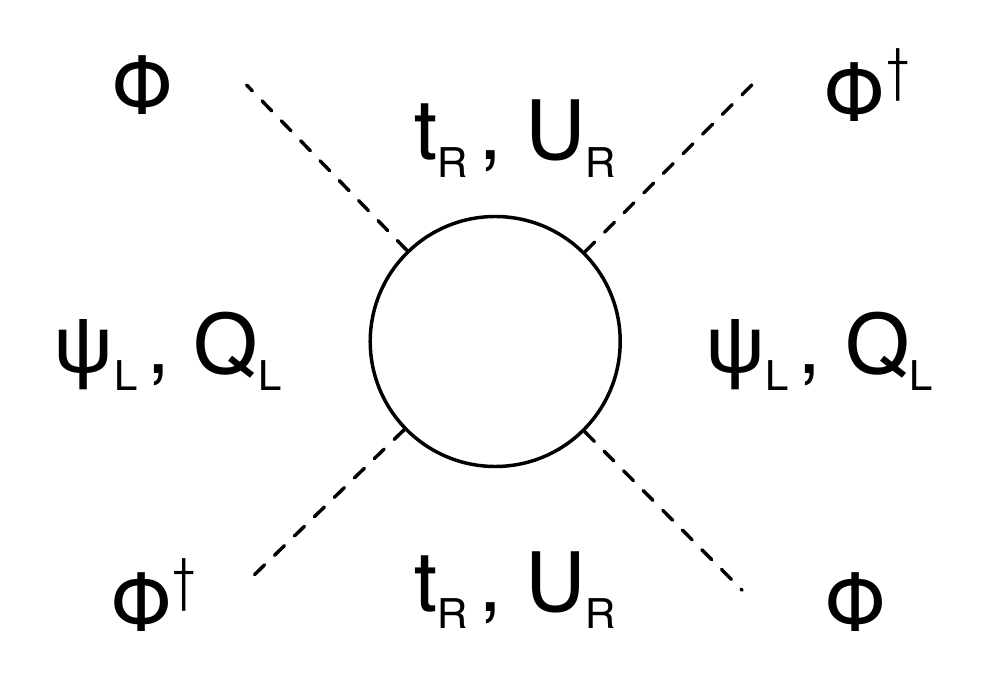}
\caption{\label{fig:FermionBubble}Diagrammatic representation of ``fermion bubble" approximation yielding
the kinetic energy and mass (left) and self-couplings (right) of the composite $\Phi$ field. The two-point function is resumed to
generate the kinetic energy term for the composite scalar field.}
\end{figure}

In the ``fermion bubble" approximation \cite{Nambu:1961tp,Bardeen:1989ds,Chivukula:1998uf}  illustrated in Fig. \ref{fig:FermionBubble}, and close to the
critical point for chiral symmetry breaking, the auxiliary field $\Phi$ becomes a light propagating composite
state. To leading order in the number of fermions (colors for quarks or technicolors for technifermions), the effects of the
strong topcolor interactions at a scale $\mu \ll \Lambda$ may be summarized by the effective Lagrangian
\begin{eqnarray}\label{eq:EFLagTopColorTC}
\mathcal{L}_{tc}&=&
D_{\mu}\Phi{D}^{\mu}\Phi-\tilde{m}^2_{\Phi}\Phi^{\dag}\Phi-\tilde{g}_t(\bar\psi_{L0}t_{R2}\Phi+h.c.)-\frac{\tilde{\lambda}}{2}(\Phi^{\dag}\Phi)^2\nonumber\\
&&-\eta\tilde{g}_t(\bar{Q}_{L}U_{R}\Phi+h.c.)-\frac{\eta^2g_t^2}{\Lambda^2}(\bar{Q}_{L}U_{R})(\bar{U}_{R}Q_L)
\end{eqnarray}
with the couplings
\begin{eqnarray}\label{eq:TopColorRenCoupling}
\tilde{g}_t^2(\mu)&=&\frac{(4\pi)^2}{(N_C+\eta^2{N_{TC}})\ln(\Lambda^2/\mu^2)},\\
\tilde\lambda(\mu)&=&2\frac{(4\pi)^2}{(N_C+\eta^2{N_{TC}})\ln(\Lambda^2/\mu^2)}.
\label{eq:TopHiggsSelfCoupling}
\end{eqnarray}
Here, in order to have a conventional kinetic energy term, we have rescaled the field $\Phi$ by
\begin{equation}
Z^{1/2}_{\Phi}=\left(\frac{g_t^2}{(4\pi)^2}(N_C+\eta^2{N_{TC}})\ln\frac{\Lambda^2}{\mu^2}\right)^{1/2}~.
\end{equation}
The mass parameter for the composite field $\Phi$ is given by,
\begin{equation}
\tilde{m}^2_{\Phi}=Z_{\Phi}^{-1}\left[\Lambda^2-\frac{2g_t^2}{(4\pi)^2}(N_C+\eta^2{N_{TC}})(\Lambda^2-\mu^2)\right]~.
\label{eq:PhiMass}
\end{equation}
The composite Higgs is light, and the effective theory valid, when $\mu \ll \Lambda$ and $g_t$ is close to the critical coupling 
$g^*_t$ for topcolor chiral symmetry given by
\begin{equation}
\frac{2{g^*_t}^2}{(4\pi)^2}(N_C+\eta^2{N_{TC}})=1~.
\label{eq:critical-coupling}
\end{equation}

For convenience, we conclude this section with a brief discussion of the $\eta \to 0$ limit. As we will see, $\eta$ will
be rather small and many of the parametric estimates that follow will derive from this limit. If we define $f$ as
the expectation value of $\Phi$ through
\begin{equation}
\langle \Phi \rangle = \begin{pmatrix}
\frac{f}{\sqrt{2}}\\
0
\end{pmatrix}~,
\label{eq:PhiVEV}
\end{equation}
we see from Eqs. (\ref{eq:EFLagTopColorTC}) and (\ref{eq:TopColorRenCoupling}) that 
\begin{equation}
m^2_t = \frac{\tilde{g}^2_t (m_t)f^2}{2} = 
\frac{(4\pi)^2 f^2}{2 N_C\ln(\Lambda^2/m_t^2)}~,
\label{eq:mtop}
\end{equation}
where we choose $\mu=m_t$ as appropriate in evaluating the top-quark mass.
This expression is usually rewritten as
\begin{equation}
f^2 = \frac{2N_C m^2_t}{(4\pi)^2}\log \left(\frac{\Lambda^2}{m^2_t}\right)~,
\label{eq:PagelsStokar}
\end{equation}
and reproduces the Pagels-Stokar relation \cite{Pagels:1979hd} appropriate in this limit  \cite{Nambu:1961tp,Bardeen:1989ds,Chivukula:1998uf}. Note that, in the effective Lagrangian of Eq. (\ref{eq:EFLagTopColorTC}), the top quark receives mass only through its coupling to the composite Higgs. Therefore, to the extent that $\eta$ is small, this relation continues to be true even after including the effects of technicolor.

\subsection{Technicolor}

Next, we consider matching\footnote{In principle, if $\Lambda/\Lambda_{TC} \gg1$, we should
also include the scaling of the operators in  Eq. (\protect\ref{eq:EFLagTopColorTC}) due to
the technicolor interactions. In practice, all of the relevant corrections can be absorbed into a redefinition
of $\eta$ -- and hence will be neglected in what follows.}
the Lagrangian in Eq. (\ref{eq:EFLagTopColorTC}) to the chiral
Lagrangian valid at scales below the scale of technicolor chiral symmetry breaking, $\Lambda_{TC}$.
In what follows, we will use the Naive Dimensional Analysis (NDA) \cite{Weinberg:1978kz,Manohar:1983md,Georgi:1985hf} 
estimate $\Lambda_{TC} \simeq 4 \pi F$, where $F$ is the technicolor pion decay constant (the analog of $f_\pi \approx 93$ MeV in QCD). To keep track of the
chiral symmetry properties of the technifermion -- scalar coupling in $\mathcal{L}_{tc}$ we introduce the $2\times 2$ matrix
\begin{equation}
{\mathcal M} = \eta \tilde{g}_t\,\left( \Phi\ 0\right)~,
\end{equation}
which serves as a spurion ``transforming" as $\mathcal{M} \to L \mathcal{M}R^\dagger$ under the $SU(2)_L \times SU(2)_R$ chiral symmetries of the technifermions. The coupling of the technifermions to the field $\Phi$, then, is similar to the mass term
in QCD, and hence we expect the effective Lagrangian
\begin{eqnarray}
\mathcal{L}^{(2)}_{TC}=\frac{F^2}{4}\textrm{tr}[D_{\mu}\Sigma^{\dag}D^{\mu}\Sigma]+4\pi F^3\left(\frac{c_1 }{2}\right) \textrm{tr}[{\cal M}^{\dag}\Sigma+\Sigma^{\dag}{\cal M}]~,
\label{eq:ChiralLagrangian}
\end{eqnarray}
where $c_1$ is an unknown
chiral coefficient related to the magnitude of the technifermion condensate which, in QCD, 
is approximately 2.\footnote{More properly, the corresponding term in the QCD chiral Lagrangian gives
\begin{equation}
m^2_\pi = 4\pi f_\pi c_1 (m_u+m_d) \approx (135\,{\rm MeV})^2
\cdot\left( \frac{m_u+m_d}{8\,{\rm MeV} }\right) \cdot \left( \frac{c_1}{2}\right)~.
\end{equation}.
}
Here $\Sigma$ is the $SU(2)_L\times SU(2)_R/SU(2)_V$ nonlinear sigma model field associated with electroweak symmetry
breaking, and is to be associated with $\Sigma_{01} \Sigma_{12}$ in the triangle Moose model described
in Sec. \ref{sec:model} above.

The second term in Eq. (\ref{eq:ChiralLagrangian}) arises from the ETC coupling of the top quark
and is of particular interest since it couples the top-color and technicolor chiral symmetries -- and hence will
give rise to the top-pion masses. To analyze this term, it is convenient to rewrite $\Sigma$ in terms of
a two-component complex unimodular vector $\xi$
\begin{equation}
\Sigma = \begin{pmatrix}
\xi & -i \sigma_2 \xi^*
\end{pmatrix}
=
\begin{pmatrix}
\xi & \tilde{\xi}
\end{pmatrix}~,
\end{equation}
where
\begin{equation}\label{eq:ConstraintSigmaDoublet}
\xi\xi^{\dag}+\tilde\xi\tilde\xi^{\dag}= {\cal I}_{2\times2},\quad
\xi^{\dag}\xi=\tilde\xi^{\dag}\tilde\xi= 1,\quad
\xi^{\dag}\tilde\xi=\tilde\xi^{\dag}\xi=0~.
\end{equation}
By the usual convention, $\xi$ has the following vacuum expectation value
\begin{equation}
\langle \xi \rangle = \begin{pmatrix}
1 \\
0
\end{pmatrix}~,
\label{eq:XiVEV}
\end{equation}
in unitary gauge. With this convention the combined $\Phi - \xi$ potential is a special case of a  two-Higgs potential
(with $F  \xi$ playing the role of a second ``Higgs"), and the second term in Eq. (\ref{eq:ChiralLagrangian}) becomes
\begin{eqnarray}
F m^2_{Mix}[\Phi^{\dag}\xi+h.c.]~,
\end{eqnarray}
with mass-squared
\begin{eqnarray}
m^2_{Mix}(\Lambda_{TC})= 4 \pi F^2 \left(\frac{c_1}{2}\right) \eta \tilde{g}_t(\Lambda_{TC})~,
\label{eq:msqmix}
\end{eqnarray}
renormalized at scale $\Lambda_{TC}=4\pi F$.

At scales $\mu < \Lambda_{TC}$, the parameters $\tilde{\lambda}(\mu)$, $m^2_{Mix}(\mu)$, and $\tilde{m}^2_\Phi$  continue
to renormalize through the top-quark loop diagrams illustrated in Fig. (\ref{fig:FermionBubble}), {\it i.e.} the formulas
in Eqs. (\ref{eq:TopColorRenCoupling}), (\ref{eq:TopHiggsSelfCoupling}), and (\ref{eq:PhiMass})  continue to apply with $\eta \to 0$.
The complete effective Lagrangian at scale $\mu$ is 
\begin{eqnarray}\label{eq:EffLagTC2HighLambda}
\mathcal{L}^{(2)}_{TC2}(\mu)&=&\frac{F^2}{4}\textrm{tr}[D_{\mu}\Sigma^{\dag}D^{\mu}\Sigma]+D_{\mu}\Phi{D}^{\mu}\Phi
-\tilde{m}^2_{\Phi}(\mu)\Phi^{\dag}\Phi+F m^2_{Mix}(\mu)[\Phi^{\dag}\xi+h.c.]\nonumber\\
&&-\tilde{g}_t(\mu)(\bar\psi_{L0}t_{R2}\Phi+h.c.)-\frac{\tilde{\lambda}(\mu)}{2}(\Phi^{\dag}\Phi)^2
\end{eqnarray}
In what follows we will need the values of these parameters evaluated at low energies, $\mu \simeq m_t$. We will find that $\eta \ll 1$; hence, in the derivations below we will apply Eqs. (\ref{eq:TopColorRenCoupling}), (\ref{eq:TopHiggsSelfCoupling}), and (\ref{eq:PhiMass}) in the $\eta\to 0$ limit.

\subsection{Minimizing the Potential and the Scalar Spectrum}

We are interested in identifying the region of parameter space where topcolor and technicolor jointly yield electroweak symmetry breaking,
{\it i.e.} $\Phi$ has the vacuum expectation value shown in Eqs. (\ref{eq:PhiVEV}) and (\ref{eq:XiVEV}),  
with\footnote{Note that the value of $F$ here differs from that in the Top Triangle Moose model, Eq. (\protect\ref{eq:Ff})
since there electroweak symmetry breaking occurs collectively through the symmetry breaking
encoded through {\it both} $\Sigma_{01}$ and $\Sigma_{12}$.}
\begin{equation}
f = v \sin \omega~, \quad\quad F=v \cos \omega~,
\label{eq:JointVEV}
\end{equation}
and where $v \approx 246\, {\rm GeV}$ is the usual weak scale. We will assume that all of the low-energy mass parameters (the 
masses of all the scalars in the spectrum and the top-quark) have the same order of magnitude, and we adopt $\mu \simeq m_t$
implicitly in what follows.

The $\Phi-\xi$ potential may be written
\begin{eqnarray}\label{eq:TC2PotentialLO}
V(\Phi,\xi)&=&\frac{\tilde{\lambda}}{2}(\Phi^{\dag}\Phi)^2+ \tilde{m}^2_{\Phi}\Phi^{\dag}\Phi-Fm^2_{Mix}[\Phi^{\dag}\xi+h.c.]\nonumber\\
&=&\frac{\tilde{\lambda}}{2}\left(\Phi^{\dag}\Phi-\frac{f_{tc}^2}{2}\right)^2-F m^2_{Mix}[\Phi^{\dag}\xi+h.c.]+const.
\end{eqnarray}
where, $f_{tc}=-2\tilde{m}_{\Phi}^2/\tilde\lambda$. Requiring the minimum of the potential to occur at (\ref{eq:JointVEV}), we
see that 
\begin{eqnarray}
\left.\frac{\partial{V}}{\partial{f}}\right|_{\langle\Phi\rangle,\langle\xi\rangle}=0
\Rightarrow\frac{\tilde\lambda}{2}f(f^2-f_{tc}^2)-\sqrt{2}m^2_{Mix}F=0
\label{eq:replace}
\end{eqnarray}
Using Eq. (\ref{eq:replace}) to eliminate  $f_{tc}^2$ in favor of $f^2$ and $m^2_{Mix}$, the potential can be rewritten as
\begin{eqnarray}\label{eq:DeformV}
V(\Phi,\xi)&=&\frac{\tilde{\lambda}}{2}\left(\Phi^{\dag}\Phi-\frac{f^2}{2}+\frac{\sqrt{2}m_{Mix}^2F}{\tilde\lambda{f}}\right)^2-Fm^2_{Mix}[\Phi^{\dag}\xi+h.c.]+const.\nonumber\\
&=&\frac{\tilde{\lambda}}{2}\left(\Phi^{\dag}\Phi-\frac{f^2}{2}\right)^2+\frac{\sqrt{2}m_{Mix}^2F}{f}\left|\Phi-\frac{f}{\sqrt{2}}\xi\right|^2+const.~,
\end{eqnarray}
which is precisely the form  found in \cite{Chivukula:2011ag}. 

From this we find
\begin{equation}
M^2_{\Pi}=\sqrt{2}m_{Mix}^2\frac{v^2}{Ff}~,
\end{equation}
and, using Eqs. (\ref{eq:mtop}) and (\ref{eq:msqmix}),
\begin{equation}
M^2_\Pi = 8\pi v m_t \cdot \left(\frac{c_1}{2}\right) \cdot \eta\cdot \frac{\cos\omega}{\sin^2\omega}~.
\label{eq:MPisq}
\end{equation}
Note that this leading contribution to the top-pion masses yields {\it degenerate} charged-
and neutral-top-pions.
The same potential also yields the top-higgs mass $M_{H_t}$,
\begin{eqnarray}\label{eq:HtMassTC2}
M^2_{H_t}&=&\tilde\lambda{f^2}+\frac{\sqrt{2}m_{Mix}^2F}{f}
=4m_t^2+M_{\Pi}^2\cos^2\omega~,
\end{eqnarray}
where the form of the relation between $M_{H_t}$, $M_\Pi$, and $\cos\omega$ is fixed from the form
of the potential \cite{Chivukula:2011ag}, and the relation between $\tilde{\lambda}$ 
and $m_t$ is fixed in the NJL approximation \cite{Bardeen:1989ds}.

Note that the TC2 theory in the NJL limit is specified primarily by four parameters: $g_t$, $\Lambda$, $\eta$, and $F$. Physical
quantities will only depend on these four parameters, up to coefficients in the chiral Lagrangian (such as $c_1$) of
order 1. Using Eqs. (\ref{eq:PagelsStokar}), (\ref{eq:JointVEV}), and (\ref{eq:MPisq}), we will
trade the parameters $g_t$, $\Lambda$, $\eta$, and $F$ for $m_t$, $v$, $\sin\omega$, and $M_\Pi$.

\subsection{Constraints from \protect{$\Delta T$}}

\begin{figure}[th]
  \centering
    \includegraphics[width=6cm]{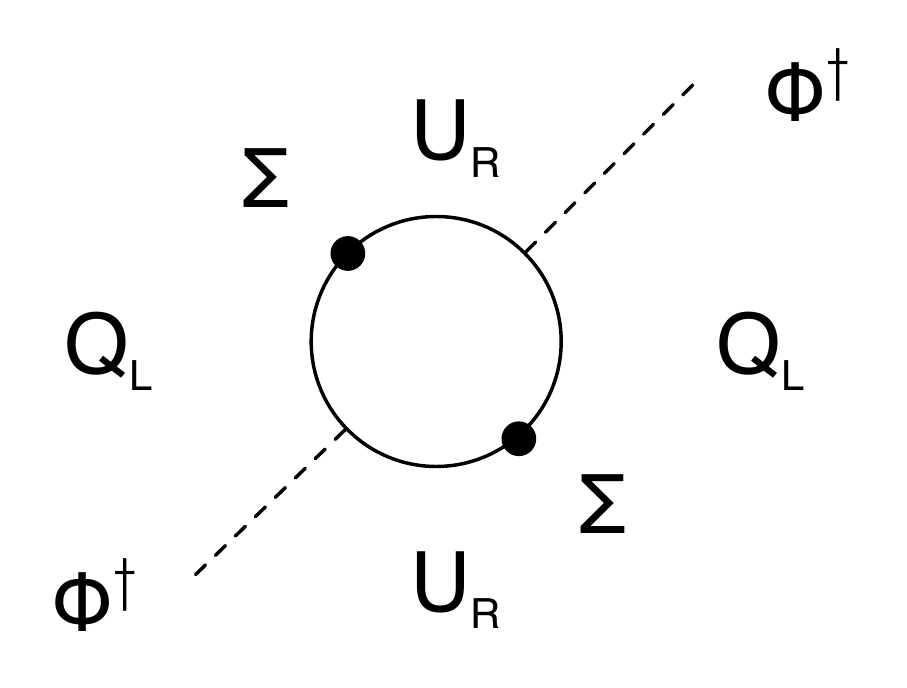}
    \hskip30pt
    \includegraphics[width=6cm]{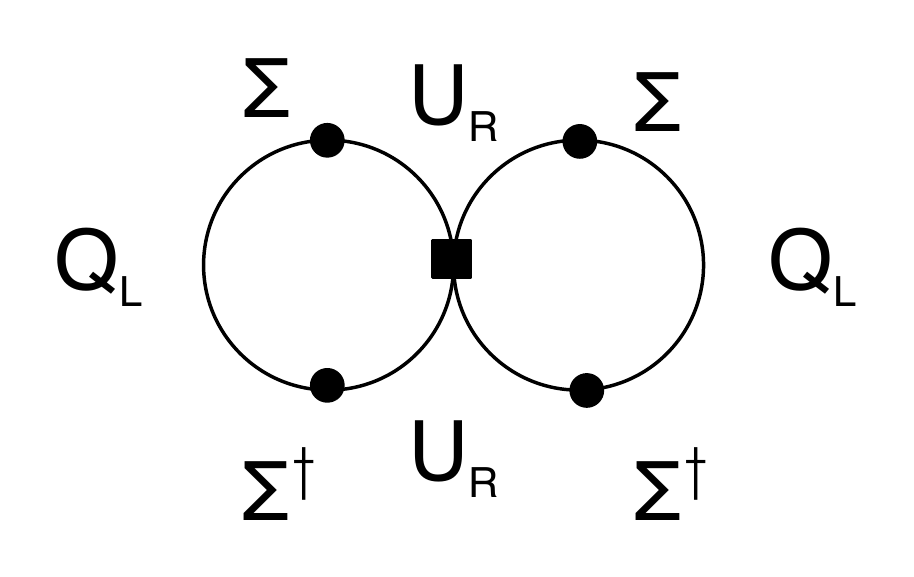}
\caption{\label{fig:T}Diagrams corresponding to the two leading contributions to
$\alpha \Delta T$ in the TC2 model. The diagram on the left gives rise to the operator shown
in Eq. (\protect\ref{eq:isospin-operator}). The diagram on the right arises from the four-technifermion operator 
shown in Eq. (\ref{eq:4fRR}). The small black circles in these diagrams represent the dynamical technifermion
mass arising from technicolor chiral symmetry breaking, as parameterized by the field $\Sigma$ in the chiral Lagrangian
of Eq. (\protect\ref{eq:ChiralLagrangian}).}
\end{figure}

The physics giving rise to the top-quark mass violates custodial isospin, causing deviations in the low-energy parameter
$\Delta \rho =\alpha \Delta T$. Consider the Lagrangian shown in Eq. (\ref{eq:EFLagTopColorTC}).
The Yukawa interaction between the composite Higgs $\Phi$ and the top-quark gives rise to the 
usual top-quark mass dependent contribution -- just as in the standard model. The last two terms in this Lagrangian,
the Yukawa interaction between the technifermions and the composite Higgs and the four-technifermion operator,
give rise to new contributions which we consider below.

\begin{figure}[!h]
  \centering%
    \includegraphics[width=8cm]{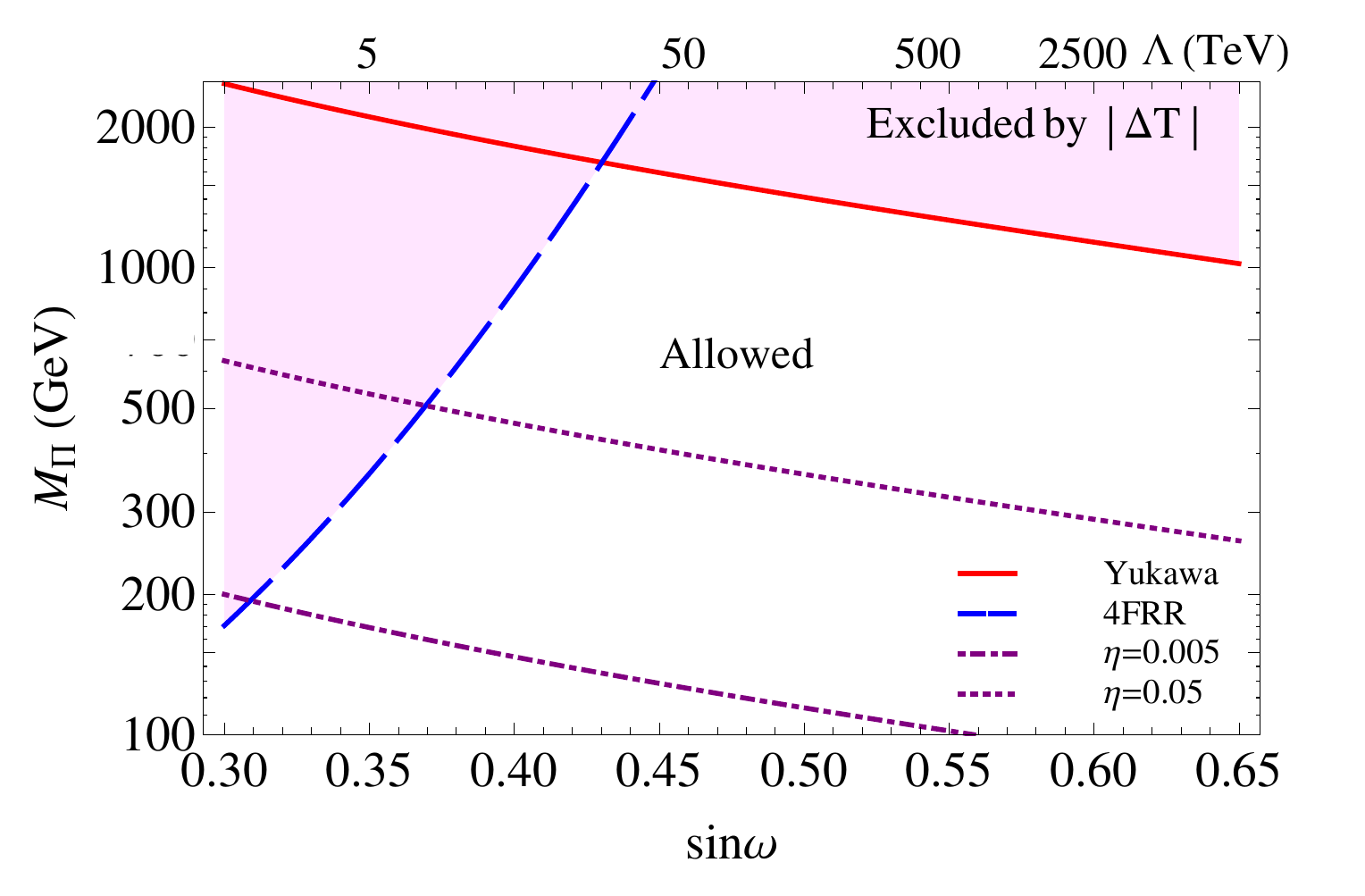}
        \includegraphics[width=8cm]{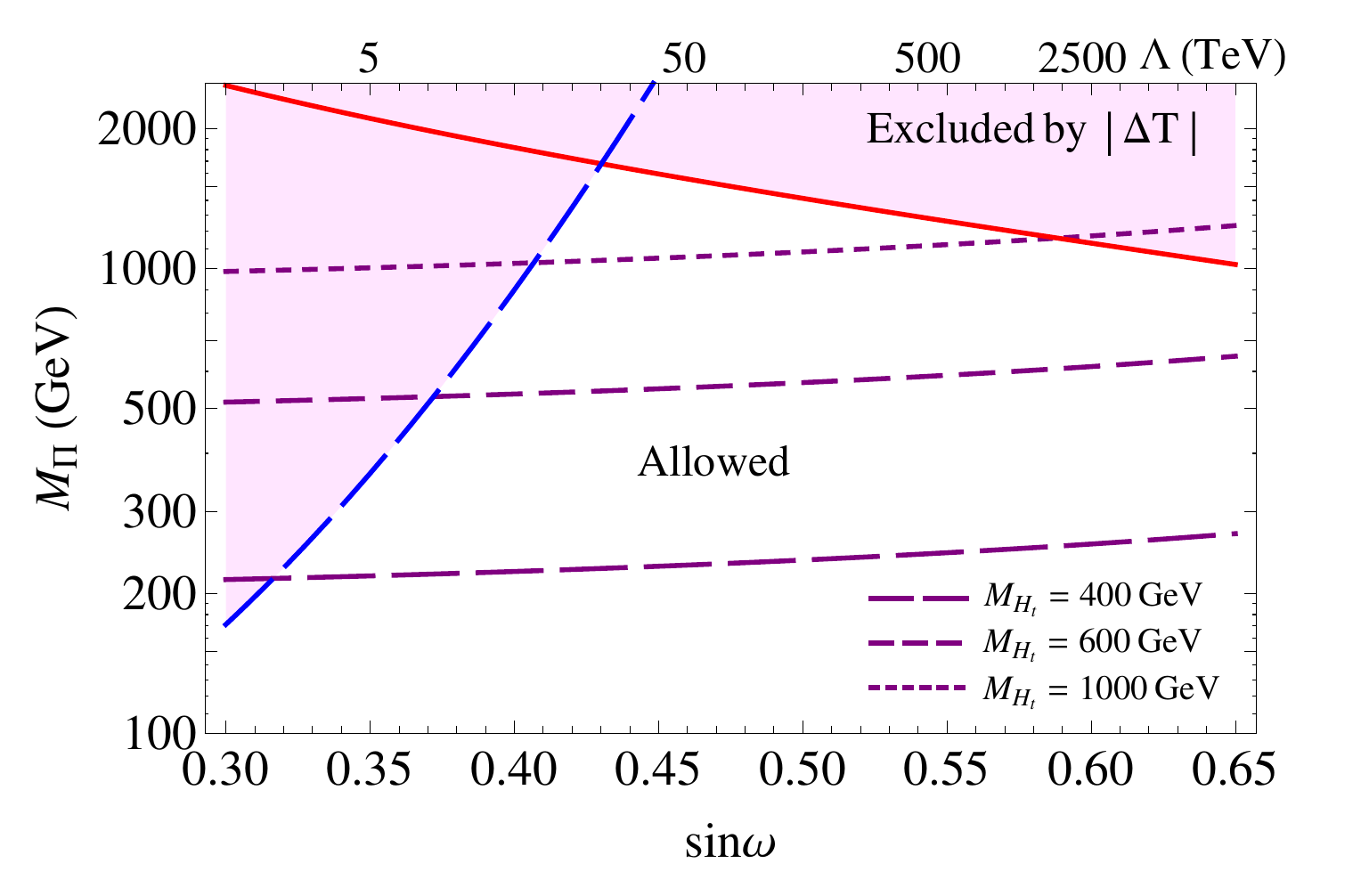}
  \caption{Left: Approximate constraints on $M_\Pi$ and $\sin\omega$ in the TC2 model in the NJL approximation coming
  from bounds on $\alpha \vert\Delta T\vert$. The constraints shown arise from taking $\vert \Delta T \vert <$ 0.5 and assuming that $c_1/2 = c_T = c_{T'} = 1$; the shaded pink region is excluded.  The red solid line shows the bound arising from the operator in Eq. (\ref{eq:Dim6})  (red line); the blue long-dashed line shows the bound from Eq. (\ref{eq:TPara4fRR})(blue dashed line).  The dotted purple curves on the left depict contours of constant $\eta$ from Eq.(\ref{eq:EFLagTopColorTC}); the dashed purple curves at right are contours of constant top-Higgs mass from Eq. (\protect\ref{eq:HtMassTC2}).}  
  \label{fig:TparaSumTC2}
\end{figure}

Consider first the technifermion Yukawa coupling. This operator violates custodial isospin by one unit, $\Delta I=1$,
and therefore the leading contribution to $\alpha \Delta T$ arises through {\it two} insertions of this operator as
shown in left-hand panel of Fig. \ref{fig:T}. This diagram yields an operator of the form
\begin{eqnarray}
&&\frac{c_{T}}{(4\pi)^2}\textrm{tr}[{\cal M}^{\dag}(D^{\mu}\Sigma){\cal M}^{\dag}(D_{\mu}\Sigma)]~,
\label{eq:isospin-operator}
\end{eqnarray}
where, consistent with NDA \cite{Weinberg:1978kz,Manohar:1983md,Georgi:1985hf}, the constant
$c_{T}$ is expected to be of order 1. Computing the effect of this operator on the $W$ and $Z$ masses, we find
\begin{equation}
\alpha\vert \Delta T\vert = \frac{2 \vert c_T\vert \eta^2 m^2_t}{(4\pi v)^2}~,
\label{eq:eta-bound}
\end{equation}
or, alternatively, rewriting
the dependence on $\eta$ in terms of $M^2_\Pi$, we find
\begin{equation}
\alpha \vert\Delta{T}\vert=\frac{\vert c_T\vert}{2} \cdot \left(\frac{2}{\vert c_1\vert}\right)^2\cdot \frac{1}{\cos^2\omega}\left(\frac{M_{\Pi}\sin\omega}{4\pi{v}}\right)^4~.
\label{eq:Dim6}
\end{equation}
If we require $\vert\Delta T \vert\lesssim0.5$, we find from Eq. (\ref{eq:eta-bound}) that $\eta \lesssim 0.6$.  The equivalent constraint, in terms of $M_\Pi$, from
Eq. (\ref{eq:Dim6}) is shown as the red solid line in  Fig. \ref{fig:TparaSumTC2}.  This is a rather weak upper bound, phenomenologically speaking.  Theoretically, it is still an interesting bound because it derives directly from the Yukawa coupling operator in the low-energy chiral expansion that also gives rise to $M_\Pi$ withough any dependence on the details of technicolor dynamics at high energies.

On the other hand, since the  ETC interaction between the top quark and technifermions in Eq. (\ref{eq:TC2ETC4fermion}) couples to both the left-handed current $\bar\psi_{L0}\gamma^{\mu}Q_L$ and right-handed current $\bar{t}_{R2}\gamma^{\mu}U_R$, it is natural to expect that
there are ETC gauge bosons that couple to $U_R$ with the same strength. The exchange of such an ETC
boson will give rise to the  $\Delta{I}=2$ operator,
\begin{eqnarray}\label{eq:4fRR}
\frac{\eta{g}_t^2}{\Lambda^2}(\bar{U}_R\gamma^{\mu}U_R)(\bar{U}_R\gamma_{\mu}U_R)~,
\end{eqnarray}
which can contribute directly to $\Delta{T}$. In particular, the diagram on the right of Fig. \ref{fig:T} yields the operator
\begin{eqnarray}
c_{T'}\cdot \frac{\eta g^2_t}{\Lambda^2} \cdot F^4 \left({\rm Tr}\left[ \Sigma^\dagger D_\mu \Sigma
\begin{pmatrix} 1 & 0 \\ 0 & 0 \end{pmatrix}
\right]\right)^2~,
\end{eqnarray}
where $c_{T'}$ is an unknown chiral coefficient of order 1.\footnote{In fact, it is exactly equal to 1 
in the vacuum insertion approximation.}
The correction to $\vert\Delta T\vert$ is
\begin{eqnarray}\label{eq:TPara4fRR}
\alpha \vert\Delta{T}\vert= \frac{4\vert c_{T'}\vert}{v^2} \cdot \frac{\eta g^2_t F^4}{\Lambda^2}~.
\end{eqnarray}
To evaluate this expression, we use Eq. (\ref{eq:MPisq}) to rephrase $\eta$ in terms of $M_\Pi$, apply  Eq. (\ref{eq:JointVEV}
) to eliminate $F$, and
approximate $g^2_t $ by ${g^*_t}^2$ as in Eq. (\ref{eq:critical-coupling}) [neglecting the term of order $\eta^2$]:
\begin{equation}
\alpha \vert \Delta T\vert = \frac{4\pi}{N_C}\cdot \vert c_{T'}\vert  \cdot \left(\frac{2}{\vert c_1\vert }\right)\cdot \sin^2\omega \cos^3\omega \cdot 
\frac{v M^2_\Pi}{m_t \Lambda^2}~.
\end{equation}
This constraint is represented by the blue long-dashed line in Fig. \ref{fig:TparaSumTC2}.

Figure \ref{fig:TparaSumTC2} summarizes the approximate constraints on the $\sin\omega - M_\Pi$ plane that arise from limits on $\alpha \vert\Delta T\vert$ as discussed above.  The pink-shaded regions are excluded; the area above the solid red line is excluded due to the impact of the technifermion Yukawa coupling and the area to the left of the blue long-dashed line is excluded by the effects of ETC gauge boson exchange.  In the left-hand pane, a few dotted curves for different values of $\eta$ are shown to indicate how that dimensionless parameter varies with $\sin\omega$ and $M_\Pi$; in the right-hand pane, a few nearly-horizontal purple contours corresponding to several values of the top-Higgs mass are shown.  


\subsection{Top-Pion Mass Splitting}

Finally, we consider the mass splitting between the charged and neutral top-pions. The leading contribution comes
from the same diagram that produces the operator in Eq. (\ref{eq:isospin-operator}). In particular, in addition to the
derivative operator discussed above, these diagrams give rise to the operators
\begin{equation}
\tilde\lambda_4F^2 \Phi^{\dag}\xi\xi^{\dag}\Phi
+\tilde\lambda_{5'}F^2 \left(\Phi^{\dag}\xi\Phi^{\dag}\xi+\xi^{\dag}\Phi\xi^{\dag}\Phi\right)
\label{eq:two-Higgs}
\end{equation}
where, using NDA, the parameters
$\lambda_i$ are
\begin{equation}
\tilde{\lambda}_i = c_i (\eta \tilde{g}_t)^2~,
\end{equation}
and the $c_i$ are parameters of order 1. Comparing the operators in Eq. (\ref{eq:two-Higgs}) with those in the two-Higgs
doublet model ($\xi$ transforms precisely as a Higgs, but with fixed magnitude) we see that these terms each
give rise to mass splittings of order
\begin{equation}
\Delta M^2_\Pi = M^2_{\Pi^+}-M^2_{\Pi^0} \propto \tilde{\lambda}_i v^2~.
\end{equation}
From the relations derived previously, we find
\begin{equation}
\Delta M^2_\Pi \propto c_i \left(\frac{2}{c_1}\right)^2\cdot \frac{M^4_\Pi}{32\pi^2 v^2} \cdot \frac{\sin^2\omega}{\cos^2\omega}~,
\end{equation}
and therefore, ignoring factors of order one
\begin{equation}
\frac{\Delta M_\Pi}{M_\Pi} \propto \left(\frac{M_\Pi}{6.2\,{\rm TeV}}\right)^2\cdot \frac{\sin^2\omega}{\cos^2\omega}~.
\end{equation}
From this we see that, for the allowed range of $M_\Pi$, the mass-splitting between the charged top-pion and the neutral top-pion is
typically very small, and always less than {\cal O}(10\%).  For $M_{\Pi_t}$ of order 200 GeV, the mass splitting is of order 100 MeV.

Based on this analysis, it is clear that the classic TC2 dynamics do not lead to large splittings between the top and neutral top-pions.  A model with a large splitting must contain additional isospin violation, beyond the minimum required to generate the top quark's mass.

\section{Alternative Fermion Couplings and Constraints from $b \to s \gamma$}
\label{sec:flavor}

The couplings of the top-pion and top-Higgs to fermions are model dependent.  In this appendix we discuss the relation between the assumptions about the flavor structure that are used in this paper and the simpler form of the fermion couplings used in \cite{Chivukula:2009ck}.

The form for the light fermion masses given in  \cite{Chivukula:2009ck} is
\begin{eqnarray}
\mathcal{L} & = & M_{D}\left[\epsilon_{L}\bar{\psi}_{L0}\Sigma_{01}\psi_{R1}+\bar{\psi}_{R1}\psi_{L1}+\bar{\psi}_{L1}\Sigma_{12}\left(\begin{array}{cc}
\epsilon_{uR} & 0\\
0 & \epsilon_{dR}\end{array}\right)\left(\begin{array}{c}
u_{R2}\\
d_{R2}\end{array}\right)\right].
\label{eqn:Light fermion mass}
\end{eqnarray}
We have denoted the Dirac mass that sets the scale of the heavy fermion masses as $M_D$.  Here, $\epsilon_{L}$ is a parameter that describes the degree of delocalization of the left handed fermions and is assumed to be universal for
the light quark generations and the leptons. All the flavor violation for the light fermions is then encoded in the last term; the delocalization parameters for the right handed fermions, $\epsilon_{fR}$, which can be adjusted to realize the masses and mixings of the up and down type fermions.  The mass of the top quark arises from similar terms with a unique left-handed delocalization parameter $\epsilon_{tL}$ and also from a unique Lagrangian term reflecting the coupling of the top-Higgs to the top quark:
\begin{equation}
\mathcal{L}_{top}=-\lambda_{t}\bar{\psi}_{L0}\,\Phi\, t_{R}+h.c.\label{eq:top quark mass L}
\end{equation}
%

\begin{figure}
\includegraphics[width=0.8\textwidth]{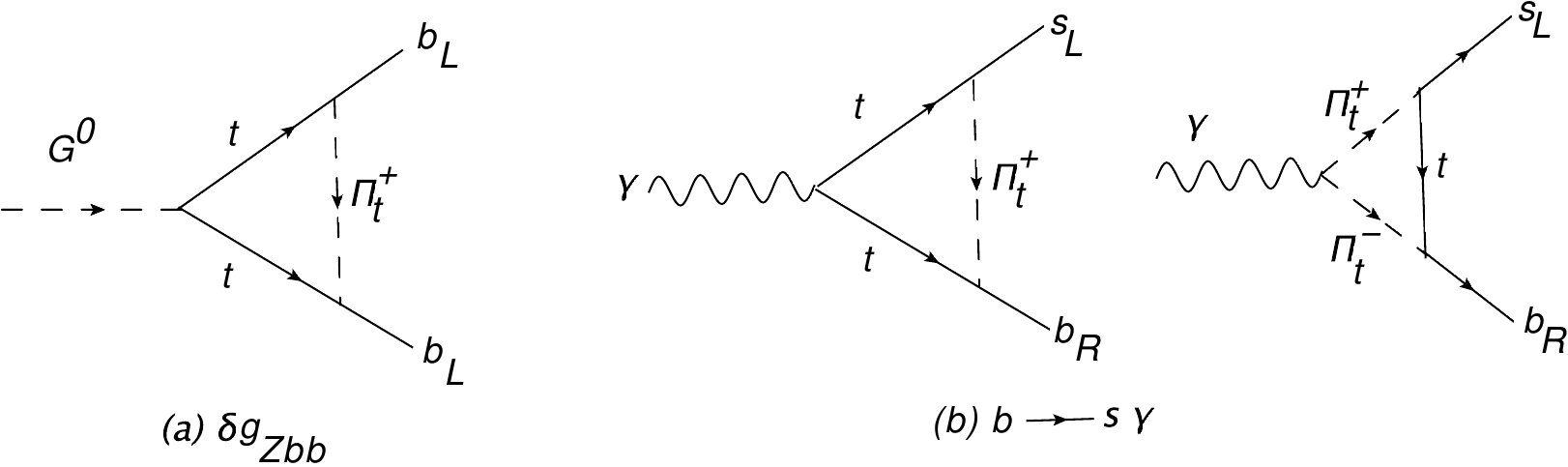}
\caption{Loop corrections to $\delta{g}_{Zbb}$ and $b\rightarrow{s\gamma}$ arising from exchange of charged top-pions.}
\label{fig:bsgamma}
\end{figure}

If this simple picture for the fermion masses is correct, then top-color provides mass only to the
top-quark while the three-site/technicolor sector provides mass to both the top-quark and all lighter-quarks.
In this case, insofar as the third-generation quarks are concerned, the pattern of top-pion couplings is
the same as the pattern of charged-Higgs couplings in ``type-II" two-Higgs-doublet models \cite{Gunion:1989we} -- with the
top-Higgs playing the role of the Higgs-doublet coupling to top-quark and the technicolor-sector playing the
role of the Higgs-doublet giving mass to the bottom. 
\begin{eqnarray}
\mathcal{L}_{yukawa}&=&(2\sqrt{2}G_F)^{1/2}\sum_{i,j}\bar{u}_i(\cot\omega\,m_{ui}V_{ij}P_L+\tan\omega\, {V_{ij}}m_{dj}P_R)d_j\Pi^++h.c.\nonumber\\
&\supset&(2\sqrt{2}G_F)^{1/2}\left[m_tV_{tb}\cot\omega\, \bar{t}_Rb_L+m_tV_{ts}\cot\omega\, \bar{t}_Rs_L+m_bV_{tb}\tan\omega\, \bar{t}_Lb_R\right]\Pi^++h.c.
\label{eq:problem}
\end{eqnarray}
These couplings imply significant corrections from charged top-pion exchange to the processes $Z \to \bar{b} b$
and $b \to s \gamma$, as illustrated in Fig. \ref{fig:bsgamma}. The correction to the process $Z \to \bar{b} b$
comes predominantly from the first term in Eq. (\ref{eq:problem}) -- and is characteristic of top color theories
\cite{Burdman:1997pf}. As explained in \cite{Chivukula:2011ag}, the top-color corrections to $Z \to \bar{b} b$
can be compensated for by
an adjustment of the top-quark delocalization parameter $\epsilon_{tL}$. 

\begin{table}[!h]
\caption{\label{tab:ChargedPionMinMassBtoSR}Lower bound on $M_{\Pi_t^+}$ from $b\rightarrow{s\gamma}$
assuming the fermion couplings in Eq. (\protect\ref{eq:problem}).}
\begin{center}
\begin{tabular}{c|c|c|c|c|c|c|c|c|c|c|c|c|c}
\hline\hline
$\sin\omega$ & 0.16 & 0.19 & 0.23 & 0.26 & 0.30 & 0.34 & 0.40 & 0.46 & 0.53 & 0.60 & 0.70 & 0.83 & 0.96\\
\hline
$M_{\Pi_t^+} (GeV)$ & 754 & 685 & 617 & 551 & 500 & 440 & 396 & 363 & 332 & 311 & 289 & 270 & 254\\
\hline\hline
\end{tabular}
\end{center}
\end{table}

The potential corrections to $b \to s \gamma$, however,
are more problematic.  These arise from vertices involving both the second interaction in Eq. (\ref{eq:problem}) [which is necessary since the process involves the strange-quark] and either the first or third one. These contributions are particularly
severe\footnote{The role of $\beta$ in type-II two-Higgs-doublet models is played here by $\omega$. In two-Higgs models
one often  considers $\tan \beta \simeq m_t/m_b \gg1$ -- while here, we are mostly interested in 
$\tan\omega =f/F \lesssim 1$.}  in the case of small $\sin\omega$. Translating
the bounds in two-Higgs models to the case at hand \cite{Deschamps:2009rh}, we find that the couplings of Eq.(\ref{eq:problem})
imply the stringent lower bounds on the charged top-pion shown in Table \ref{tab:ChargedPionMinMassBtoSR}.
Charged top-pion masses of this order, and hence neutral pion and top-Higgs masses which are expected
to be of the same order, would be very difficult to observe at the LHC.  As discussed in the text, this constraint does not apply if left-handed mixing is purely in the up-quark sector.

\section{Formulas for the top-Higgs decay widths}
\label{sec:tophiggsdecays}

The couplings of the top-Higgs, along with its decay widths to the relevant channels $WW$, $ZZ$, $t\bar{t}$, $\Pi_t^{\pm}W^{\mp}$, $\Pi_t^0Z$, $\Pi_t^{+}\Pi_t^{-}$, and $\Pi_t^0 \Pi_t^0$, were given in Ref.~\cite{Chivukula:2011dg}.  For completeness, we reproduce the key formulas below.

For the limit-setting in Sec.~\ref{sec:tophiggslimits}, we compute the top-Higgs production cross section with the aid of the 7~TeV LHC SM Higgs cross sections in the gluon fusion and VBF modes from Ref.~\cite{xsecwg}.  To the extent that the narrow-width approximation is valid, we can write
\begin{eqnarray}
	\frac{\sigma(pp \to H_t \to WW)}{\sigma(pp \to H_{\rm SM} \to WW)}
	&=& \frac{ \left[ \sigma_{gg}(pp \to H_t) + \sigma_{\rm VBF}(pp \to H_t) \right]
	{\rm BR}(H_t \to WW)}
	{ \left[ \sigma_{gg}(pp \to H_{\rm SM}) + \sigma_{\rm VBF}(pp \to H_{\rm SM}) \right]
	{\rm BR}(H_{\rm SM} \to WW)} \nonumber \\
	&\simeq& \frac{ \frac{1}{\sin^2\omega} \sigma_{gg}(pp \to H_{\rm SM})
	+ \sin^2\omega \, \sigma_{\rm VBF}(pp \to H_{\rm SM})}
	{\sigma_{gg}(pp \to H_{\rm SM}) + \sigma_{\rm VBF}(pp \to H_{\rm SM})}
	\times \frac{{\rm BR}(H_t \to WW)}{{\rm BR}(H_{\rm SM} \to WW)},
\end{eqnarray}
and analogously for the $ZZ$ final state [note that ${\rm BR}(H_t \to ZZ)/{\rm BR}(H_{\rm SM} \to ZZ)  = {\rm BR}(H_t \to WW)/{\rm BR}(H_{\rm SM} \to WW)$].  The approximation in the second line is exact insofar as (i) the QCD corrections to Higgs production are the same for the top-Higgs and the SM Higgs and (ii) the efficiencies of the inclusive LHC Higgs searches are the same for events arising from gluon fusion and VBF.

For decays to a top-pion and a gauge boson,
\begin{eqnarray}
  \Gamma(H_t \to \Pi^{\pm}_t W^{\mp}) &=& \frac{\cos^2\omega}{8 \pi v^2}
  M_{H_t}^3 \beta_W^3, \nonumber \\
  \Gamma(H_t \to \Pi^0_t Z) &=& \frac{\cos^2\omega}{16 \pi v^2}
  M_{H_t}^3 \beta_Z^3,
  \label{eq:HPiV}
\end{eqnarray}
where
\begin{equation}
  \beta_V^2 \equiv \left[ 1 - \frac{(M_{\Pi_t} + M_V)^2}{M_{H_t}^2} \right]
  \left[ 1 - \frac{(M_{\Pi_t} - M_V)^2}{M_{H_t}^2} \right].
\end{equation}
For decays to two top-pions,
\begin{eqnarray}
  \Gamma(H_t \to \Pi_t^+ \Pi_t^-) &=& 
  \frac{\lambda^2_{H\Pi^+\Pi^-}}{16 \pi M_{H_t}}
  \sqrt{1 - \frac{4 M_{\Pi_t^+}^2}{M_{H_t}^2}}, \nonumber \\
 \Gamma(H_t \to \Pi_t^0 \Pi_t^0) &=&
  \frac{\lambda^2_{H\Pi^0\Pi^0}}{32 \pi M_{H_t}}
  \sqrt{1 - \frac{4 M_{\Pi_t^0}^2}{M_{H_t}^2}},
\end{eqnarray}
where
\begin{eqnarray}
  \lambda_{H\Pi^+\Pi^-} &=& \frac{1}{v \sin\omega}
  \left[ M_{H_t}^2 \cos^2\omega - M_{\Pi_t^+}^2 + 2 M_{\Pi_t^+}^2 \sin^2\omega \right], \nonumber \\
    \lambda_{H\Pi^0\Pi^0} &=& \frac{1}{v \sin\omega}
  \left[ M_{H_t}^2 \cos^2\omega - M_{\Pi_t^+}^2 + 2 M_{\Pi_t^0}^2 \sin^2\omega \right].
\end{eqnarray}
For decays to top-quark pairs,
\begin{equation}
\Gamma(H_t \to t\bar{t}) = \frac{3  m^2_t}{8\pi v^2 \sin^2\omega} M_{H_t}
\left(1-\frac{4 m^2_t}{M^2_{H_t}}\right)^{3/2}~.
\label{eq:Httbar}
\end{equation}
By comparison, the width to gauge-bosons is suppressed
by $\sin^2\omega$:
\begin{eqnarray}
	\Gamma(H_t \to W^+W^-) &=&
	\frac{M_{H_t}^3 \sin^2\omega}{16 \pi v^2}
	\sqrt{1 - x_W} \left[ 1 - x_W + \frac{3}{4} x_W^2 \right],
	\nonumber \\
	\Gamma(H_t \to ZZ) &=&
	\frac{M_{H_t}^3 \sin^2\omega}{32 \pi v^2}
	\sqrt{1 - x_Z} \left[ 1 - x_Z + \frac{3}{4} x_Z^2 \right],
\label{eq:HVV}
\end{eqnarray}
where $x_V = 4 M_V^2 / M_{H_t}^2$.


\end{document}